\documentclass[conference]{IEEEtran}
\IEEEoverridecommandlockouts
\usepackage{cite}
\usepackage{tabularx} 
\usepackage{arydshln}
\usepackage{amsmath,amsthm,amssymb,amsfonts}
\usepackage{algorithmic}
\usepackage{graphicx}
\usepackage{textcomp}
\usepackage{xcolor}
\usepackage{balance}
\usepackage{booktabs}
\usepackage{subcaption}
\usepackage{multirow}
\usepackage{bm}
\usepackage{bbm}
\usepackage{bbding}
\usepackage{microtype}
\usepackage{clrscode}
\usepackage{multicol}
\usepackage{float}
\usepackage{color}
\usepackage{amsopn}
\usepackage{mathtools}
\usepackage{blkarray}
\usepackage{enumerate}
\usepackage{courier}
\usepackage{mathrsfs}
\usepackage{rotating}
\usepackage{wrapfig}
\usepackage{array}
\usepackage{ragged2e}
\usepackage{hyperref}
\usepackage{threeparttable}

\def\BibTeX{{\rm B\kern-.05em{\sc i\kern-.025em b}\kern-.08em
    T\kern-.1667em\lower.7ex\hbox{E}\kern-.125emX}}
\begin{document}

\title{UFIN: Universal Feature Interaction Network for Multi-Domain Click-Through Rate Prediction}

\author{\IEEEauthorblockN{Zhen Tian}
\IEEEauthorblockA{\textit{Gaoling School of Artificial Intelligence} \\
\textit{Renmin University of China}\\
Beijing, China \\
chenyuwuxinn@gmail.com}
\and
\IEEEauthorblockN{Changwang Zhang}
\IEEEauthorblockA{\textit{Poisson Lab, Huawei} \\
Beijing, China \\
changwangzhang@foxmail.com}
\and
\IEEEauthorblockN{Wayne Xin Zhao}
\IEEEauthorblockA{\textit{Gaoling School of Artificial Intelligence} \\
\textit{Renmin University of China}\\
Beijing, China \\
batmanfly@gmail.com}
\and
\IEEEauthorblockN{Xin Zhao}
\IEEEauthorblockA{\textit{Poisson Lab, Huawei} \\
Beijing, China \\
zhaoxin151@huawei.com}
\and
\IEEEauthorblockN{Ji-Rong Wen}
\IEEEauthorblockA{\textit{Gaoling School of Artificial Intelligence} \\
\textit{Renmin University of China}\\
Beijing, China \\
 jrwen@ruc.edu.cn}
\and
\IEEEauthorblockN{Zhao Cao}
\IEEEauthorblockA{\textit{Poisson Lab, Huawei} \\
Beijing, China \\
caozhao1@huawei.com}
}
\newcommand{\ie}{\emph{i.e.,} }
\newcommand{\eg}{\emph{e.g.,} }
\newcommand{\paratitle}[1]{\vspace{1.5ex}\noindent\textbf{#1}}
\newcommand{\modified}[1]{\textcolor{blue}{#1}}
\maketitle
\newtheorem{theorem}{Theorem}
\theoremstyle{theorem}
\newtheorem{mytheory}{Theorem}
\newtheorem{myproof}{Proof}
\newtheorem*{remark}{Remark}

\begin{abstract}
Click-Through Rate (CTR) prediction, which aims to estimate the probability of a user clicking on an item, is a key task in online advertising.
Numerous existing CTR models concentrate on  modeling the feature interactions within a solitary domain, thereby rendering them inadequate for fulfilling the requisites of multi-domain recommendations in real industrial scenarios.
Some recent approaches propose intricate architectures to enhance knowledge sharing and augment model training across multiple domains. 
However, these approaches encounter difficulties when being transferred to new recommendation domains, owing to their reliance on the modeling of ID features (\eg \emph{item\_id}).
To address the above issue, we propose the \underline{\textbf{U}}niversal \underline{\textbf{F}}eature \underline{\textbf{I}}nteraction \underline{\textbf{N}}etwork (\textbf{UFIN}) approach for CTR prediction.
UFIN exploits textual data to learn universal feature interactions that can be effectively transferred across diverse domains.
For learning universal feature representations, we regard the text and feature as two different \emph{modalities} and  propose an encoder-decoder network founded on a Large Language Model (LLM)  to enforce the transfer of data from the text modality to the feature modality.
Building upon the above foundation, we further develop a mixture-of-experts (MoE) enhanced adaptive feature interaction model to learn transferable collaborative patterns across multiple domains.
Furthermore,  we propose a multi-domain knowledge distillation framework to enhance  feature interaction learning.
Based on the above methods, UFIN can effectively bridge the semantic gap to learn common knowledge across various domains, surpassing the constraints of ID-based models.
Extensive experiments conducted  on eight datasets show the effectiveness of UFIN, in both multi-domain and cross-platform settings.
Our code is available at \url{https://github.com/RUCAIBox/UFIN}.
\end{abstract}


\section{Introduction}
Click-Through Rate (CTR) prediction, which aims to predict the probability of a user clicking on an item, is an important task for online advertising and recommender systems.
Various approaches have been proposed for effective CTR prediction~\cite{cheng2016wide,he2017neural,li2020interpretable,shan2016deep}. 
These methods mainly focus on accurately modeling the complicated feature
interactions to capture the underlying collaborative patterns.
Most of the existing approaches concentrate on single-domain prediction, where each model is solely trained to serve the CTR prediction of a single scenario.
However, in large-scale corporate enterprises, numerous business domains frequently necessitate CTR prediction to augment user contentment and enhance commercial revenue.
For instance, in the case of e-commerce enterprises, the advertising scenarios encompass a wide array of options and manifest notable disparities, encompassing domains such as motion pictures, literary works, electronic devices, and culinary delights, among others.
Merely mixing all the data and training a single shared CTR model cannot yield satisfactory results across all domains owing to the substantial distribution variance among diverse scenarios (\emph{domain seesaw phenomenon}~\cite{chang2023pepnet}).
The domain-specific modeling paradigm severely restricts the efficient utilization of extensive user behavior data in business scenarios.

Some recent studies~\cite{sheng2021one, chang2023pepnet, zhang2022scenario, zhou2023hinet} propose conducting  multi-domain CTR predictions.
The core idea of these approaches is to introduce a shared neural network for learning the common knowledge across diverse domains, while simultaneously integrating multiple domain-specific sub-networks to capture the distinct characteristics of each domain.
Although somewhat efficacious, the majority of these methods rely on modeling the ID features (\eg \emph{item\_id}) to develop the CTR prediction.
A major obstacle of this paradigm is the limited transferability of the learned model to new recommendation scenarios, even when the underlying data structures remain unchanged.

Inspired by recent advancements in natural language recommendations~\cite{lin2023recommender}, our objective is to devise a novel approach to learn universally applicable collaborative patterns by surpassing the constraints of ID features.
Our fundamental concept entails transforming raw features, such as the \emph{location} of an item, into textual data and employing Large Language Models (LLMs) to acquire transferable representations.
While previous attempts have demonstrated the promise of this approach for certain recommendation tasks~\cite{hou2022towards, li2023text}, there remain several critical challenges to address in the context of multi-domain CTR predictions.
First, the textual semantic space is not directly conducive to the task of CTR prediction~\cite{li2023ctrl}.
In comparison to traditional feature interaction based methods, LLMs encounter difficulties in capturing collaborative patterns, thereby resulting in suboptimal model performance.
Second, due to substantial distribution variance across different domains, effectively leveraging the collaborative knowledge from the source domain to enhance the target domain proves to be a formidable task.
For instance, the interaction between features \emph{user} and \emph{title} proves most valuable in movie recommendations, yet its efficacy diminishes in the context of beauty recommendations.

To address these issues, in this paper, we propose the \underline{\textbf{U}}niversal \underline{\textbf{F}}eature \underline{\textbf{I}}nteraction \underline{\textbf{N}}etwork  (\textbf{UFIN}) for multi-domain CTR prediction.
UFIN exploits textual data to acquire knowledge of universal feature interactions that can be effectively transferred across diverse domains.
To learn universal feature representations, we devise a prompt to convert the raw features into text and subsequently generate a set of \emph{universal features} to capture the general attributes of interactions.
Notably, we regard the text and feature representations as two \emph{modalities} and devise an encoder-decoder network founded on a Large Language Model (LLM) to enforce the conversion of data from the text modality to the feature modality.
This scheme can be denoted as "raw features $\Rightarrow$ text $\Rightarrow$ universal features".
For learning universal feature interactions, we develop an MoE-enhanced adaptive feature interaction model, which can learn the generalized collaborative patterns from diverse domains.
To further enhance the acquisition of  collaborative knowledge, we propose a multi-domain knowledge distillation framework to supervise the training of our approach.
Through these aforementioned mechanisms, UFIN can effectively bridge the semantic gap to learn common knowledge across various recommendation domains, surpassing the limitations of ID-based models.

The paper's main contributions are summarized as follows:

$\bullet$ 
We propose a novel Universal Feature Interaction Network (UFIN) for CTR prediction, intelligently acquiring the collaborative patterns across diverse domains. 

$\bullet$ 
To the best of our knowledge, UFIN is the first deep CTR model to harness Large Language Models (LLMs) to adaptively learn the feature interactions for recommendations, thereby obtaining universal feature representations from textual data. This empowers UFIN to proficiently bridge the semantic gap across various domains.

$\bullet$ 
We propose a multi-domain knowledge distillation framework for enhancing the feature interaction learning.
This motivates UFIN to proficiently acquire the collaborative knowledge from diverse domains, thereby improving the model performance.

$\bullet$ We conduct extensive experiments on eight widely used datasets.
UFIN outperforms a number of competitive baselines in both multi-domain and cross-platform settings, demonstrating the effectiveness of our approach.




\section{Methodology}
In this section, we present a universal feature interaction network for multi-domain CTR predictions, named \textbf{UFIN}.
Unlike previous works, our goal is to learn the universal feature interactions that are  able to effectively transferred to new recommendation domains.

\begin{figure*}[h]
    \centering
    \includegraphics[width = 1.\textwidth]{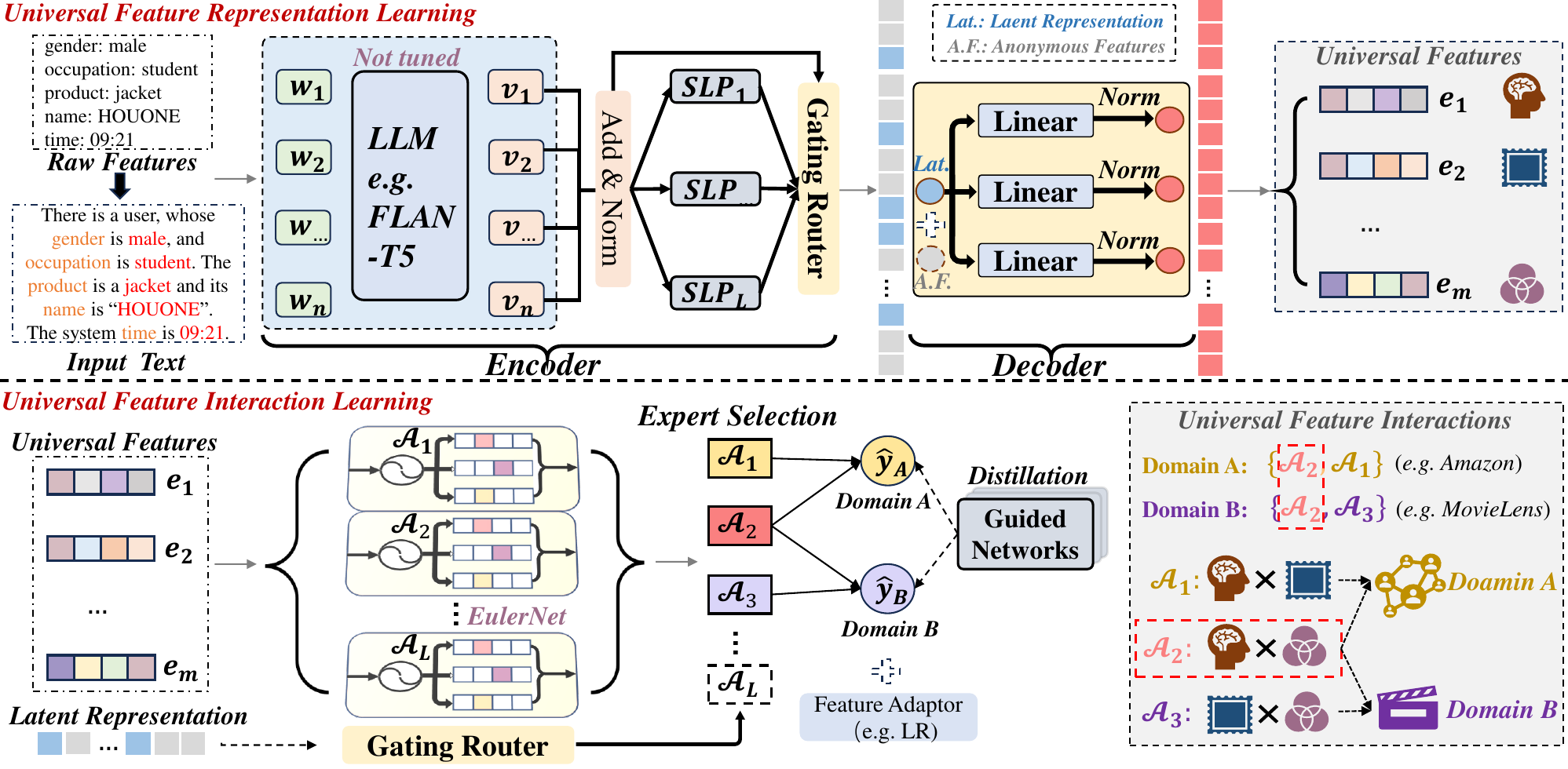}
    \caption{The overall framework of UFIN is designed as an encoder-decoder architecture, followed by a feature interaction network.	
The encoder-decoder accomplishes the transformation from text to feature modality, yielding a collection of universal features. 
The feature interaction network consists of multiple adaptive learning experts, and each expert automatically learns the underlying true interaction orders within each given domain.
Further, a semantic gating router is incorporated to adaptively integrate all experts  for learning the \emph{universal feature interactions} across diverse domains.
    }
    \label{fig:framework}
\end{figure*}

\subsection{Problem Formulation}

The CTR prediction task is to estimate the probability of a user clicking on an item.
An instance of the  CTR task can be denoted as a tuple $(\bm x, y)$, where the vector $\bm x$ includes the user, item and context features, and the label $y \in \{0, 1\}$ represents the whether the item is clicked or not.
The feature vector $\bm X$ contains multiple fields $\{ \bm x_1,  \bm x_2,..., \bm x_m \}$, where $m$ is the number of feature fields, and $\bm x_j$ is a one-hot identifier (ID) vector of the $j$-th feature field.
In this way, the CTR dataset of a single domain $D$ can be formulated as a set of \emph{instances} $\{(\bm X_i, y_i)\}_{i = 1}^N$.
Real-world recommender applications often have to deal with multiple business domains.
Specifically, multi-domain CTR models need to make predictions for $L$ domains, $\{D_1, D_2, \cdots, D_L\}$.
Compared to single-domain CTR prediction, multi-domain CTR prediction often poses greater challenges, as different domains may have different feature fields and data distributions.

Most of multi-domain CTR prediction approaches adopt the embedding-based feature interaction learning paradigm.
Since the input feature is usually sparse and high-dimensional, most CTR models employ an embedding layer to map these one-hot features into low-dimensional vectors, \ie for the feature vector $\bm x_j$, the corresponding embedding $\bm e_j$ is obtained by an embedding look-up operation.
As such, the features can be represented as a list of embeddings $ \bm E = \{ \bm{e}_1,\bm{e}_2,..,\bm{e}_m \}$.
Based on $\bm E$, most approaches conduct feature interactions $\mathcal{F}$ to capture the collaborative patterns between users and items. 
The main goal of the CTR prediction is to learn the prediction function $\hat{y} = \mathcal{F}(\bm E)$.

\subsection{Overview of UFIN}
As shown in Figure~\ref{fig:framework}, UFIN is designed with an encoder-decoder architecture followed by an interaction network.	
The core idea of UFIN is to consider the text and features as two \emph{modalities} for learning universal feature representations and universal feature interactions.	
Based on this idea, we formulate the general feature form as the natural language text and employ an MoE-enhanced large language model (LLM) as the encoder to transform the text modality into a latent space.	
Afterwards, the decoder performs a mapping to the feature modality, generating the \emph{universal features}.	
As such, UFIN can capture the generalized attributes of interactions, bridging the semantic gap between different domains.	
For learning universal feature interactions, we derive adaptive feature interactions based on the universal features to acquire transferable collaborative knowledge.
To learn the common collaborative patterns of different domains, we incorporate them into an MoE model to enhance the knowledge sharing.
Furthermore, we propose a multi-domain knowledge distillation framework to enhance feature interaction learning.	

In what follows, we introduce the details of universal feature representation learning (Section~\ref{sec:rep}) and universal feature interaction learning (Section~\ref{sec:ufi}).

\subsection{Universal Feature Representation Learning} \label{sec:rep}
To deliver transferable recommendations, we first map the features from various domains to  a common semantic space.
Prior studies rely on the embedding look-up operation to acquire the representations for the given feature IDs.	
These approaches have two significant limitations.
Firstly, it relinquishes the innate semantic information of features, thereby greatly impairing the model's transferability. 
Secondly, these models cannot be transferred to alternative platforms that possess distinct feature fields (for instance, transitioning from \emph{Amazon} to \emph{MovieLens}).

To learn transferable feature representations, we adopt natural language text as the universal data form, which is derived from raw features through a prompt.	
As increasingly more evidence shows~\cite{li2023ctrl}, text and feature representations can be regarded as two \emph{modalities} that can be mutually transformed.
Based on this idea, we employ an MoE-enhanced LLM as the encoder to enforce the text modality into a latent space and develop a decoder that performs a mapping to the feature modality, thereby generating the universal feature representations (\emph{universal features}).
This approach can be expressed as "raw features $\Rightarrow$ text $\Rightarrow$ universal features", 
providing a means to bridge the semantic gap across different domains.

\subsubsection{Feature Textualization}


The first step in learning universal feature representations is to transform the raw features into textual data, as described by a prompt.	
As previous work shows~\cite{cui2022m6}, an effective prompt should consist of \emph{personalized} fields for different users and items.
For this purpose, we design a prompt, that  includes the user profile, item description and contextual information to conduct such transformation.
As shown in Figure~\ref{fig:framework}, given the user features (\ie \emph{male}, \emph{student}), item features (\ie \emph{jacket}, \emph{HOUONE}) and context features (\ie \emph{09:21}), the transformed textual data is shown as:
\begin{quote}
There is a user, whose gender is male, and occupation is student. The product is a jacket and its name is “HOUONE”. The system time is 09:21.
\end{quote}

In our prompt, different types (\ie user-side, item-side and context-side) of raw features (\ie $\{ \bm x_1,  \bm x_2,..., \bm x_m \}$) are sequentially summarized into a natural language sentence, where the descriptions of different sides are separated by the period \verb|"."|, and the features are separated by the comma \verb|","|.
As such, a CTR instance can be denoted as $\{ \bm w_1,  \bm w_2,..., \bm w_n \}$, where $n$ is the number of words in a sentence.
In this way, we can obtain a universal data form (\ie natural language text) to represent the CTR instances across various  domains or platforms.
Note that we can also design other types of prompts to conduct such transformation. 
Unlike prior work~\cite{li2023ctrl}, the format of the prompt does not have an obvious effect in our framework, while the semantic information of the prompt, specifically the features it contains, has a large influence on the performance, which will be explored in the Section~\ref{sec:prompt}.

\subsubsection{Textual Feature Encoding}
Given the universal data form in the text modality, our goal is to transform it  into feature modality to obtain the universal features.
Most multimodal  works~\cite{radford2021learning} (\eg text-to-image)  employ an encoder-decoder architecture to align the representations of different modalities.
Following them, we first employ an MoE-enhanced LLM as the encoder to project the textual data into a common latent space.

\paratitle{LLM based Textual Encoding.}
Motivated by the recent advances  in large language models (LLMs~\cite{zhao2023survey}), which show excellent language modeling capacity, we adopt FLAN-T5~\cite{chung2022scaling} to learn the latent representations of the text.
Given the words of textual data $\{\bm w_1, \bm w_2, ..., \bm w_n\}$, we feed them into the LLM, and we have:
\begin{align}
    \{\bm v_1, \bm v_2, ..., \bm v_n\} &= \textrm{LLM}(\{\bm w_1, \bm w_2, ..., \bm w_n\}),\\
    \bm s &= \textrm{LayerNorm}(\sum_{j = 1}^n \bm v_j),
\end{align}
where $\{\bm v_1, ..., \bm v_n\} \in \mathbb{R}^{d_V \times n}$ is the last hidden state of LLM, $d_V$ is the state dimension and $\bm s$ is the latent representation.
Unlike existing studies~\cite{hou2022towards, li2023text}, we employ sum pooling to preserve token-level semantics of features and apply \textrm{LayerNorm}~\cite{ba2016layer} to adjust the semantic distributions.
\textbf{Note that the LLM is solely  for text encoding, which is not tuned during training}. 
Therefore, we can cache the last hidden state $\{\bm v_1, ..., \bm v_n\}$ to ensure the efficiency of our approach.

\paratitle{Multi-Domain Semantic Fusion.}
In the above, we obtain semantic representations from LLMs.
However, recent study~\cite{hou2022towards} found that the original semantic space of PLMs is not suitable for the recommendation task. 
To address this issue, a commonly used approach is to employ a neural network to learn the appropriate semantic space for enhancing the representations.
Since different domains usually correspond to varying semantic contexts, merely learning a shared semantic space for all domains will suffer from the domain seesaw phenomenon that degrades the model capacity.	
Inspired by the recent study~\cite{hou2022towards}, our idea is to learn an independent semantic space for each domain and adaptively combine them based on the semantic context.	

To achieve this purpose, we employ a single-layer perception (SLP) to learn a suitable semantic space for each domain, and incorporate them into an MoE model to enhance the domain fusion.
Specifically, given $L$ domains, we introduce $L$ experts, each learning in a different subspaces, and combine them through a gating router:
\begin{align}\label{eq:ec-moe}
    \bm z &= \sum_{j = 1}^L \sigma(\bm W_j \bm s + \bm b_j) \cdot g_j,\\
    \bm g &= \mathrm{Softmax}(\bm W_g \bm s)\label{eq:gate},
\end{align}
where $\bm W_j \in \mathbb{R}^{d_V \times d_V}$ and $\bm b_j \in \mathbb{R}^{d_V}$ are the weight and bias of the $j$-th expert, $\sigma$ is the activation function, $g_j$ is the $j$-th combination weight of the gating router calculated by Eq.~\eqref{eq:gate}, $\bm W_g \in \mathbb{R}^{L \times d_V}$ is the router weight, and $\bm z$ is the enhanced representations.

\paratitle{Fusing Anonymous Features.}\label{sec: faf}
As there exists numerous anonymous features (such as the identifiers of the user/item) that lack semantic information, we refrain from including them in our prompt template. Nevertheless, these features may play a significant role, particularly in situations where semantic features are limited or nonexistent. For instance, in the case of the Amazon dataset, only \emph{user\_id} features are accessible on the user side. To generate more comprehensive predictions, we expand our methodology to encompass these features. 
There are many ways to achieve this purpose, and we follow the existing ID-Text fusion work~\cite{hou2022towards}, which employs a distinct embedding for each anonymous feature and merges them with the textual representations:
\begin{align}
    \tilde{\bm z} &= \bm z + \sum_{k = 1}^{c} \bm U_k \bm h_k,\label{eq:lat}
\end{align}
where $\{\bm h_1, ..., \bm h_c\} \in \mathbb{R}^{d_A \times c}$ represents the anonymous embeddings, $c$ is the number of anonymous fields, and $\bm U_k \in \mathbb{R}^{d_V \times d_A}$ is the projection matrix. \textbf{Note that the anonymous features are only auxiliary representations and are not used unless specified.}
For efficiency considerations, we do not employ other complex mechanisms (e.g., self-attention~\cite{NIPS2017_3f5ee243}), which will be studied in our future work.


\subsubsection{Universal Feature Generation}\label{sec:uni}
With the above textual encoding procedure, we can obtain the universal representation of the instances.
Previous works~\cite{lin2021m6} directly feed the textual representations into a feedforward network (\eg MLP) to make predictions, resulting in suboptimal performance.	
As the recent study shows~\cite{rendle2020neural}, it is challenging for an MLP to capture effective collaborative patterns compared with the feature-wise interactions (\eg FM~\cite{rendle2010factorization}).
Our proposed solution, by contrast, entails harnessing textual data to generate \emph{universal features} that transcend various domains, thereby capturing the collective patterns that are commonly observed.
To illustrate, we anticipate generating the universal attribute \emph{"amusing"} from the textual expressions \emph{"This movie is hilarious"} and \emph{"This book is whimsical"} within the realms of movie and book recommendations.
Based on this idea, we develop a decoder that conducts a transformation to map the latent representation of textual data to the feature modality, shown as:
\begin{align}
    \tilde{\bm e}_j &= \textrm{LayerNorm}(\bm V_j \tilde{\bm z})\label{eq:unf},
\end{align}
where $j \in \{1, 2, ..., n_u\}$ and $n_u$ is the field number of universal features.
Here we incorporate a set of projection matrices
$\{\bm V_j \in \mathbb{R}^{d \times d_V}\}_{j = 1}^{n_u}$ to generate a set of universal features $\tilde{\bm E} = \{\tilde{\bm e}_1, \tilde{\bm e}_2, ..., \tilde{\bm e}_{n_u}\}$, each measuring different aspects from different representation subspaces.
As such, we use the generated universal representations for subsequent feature interaction modeling.

\subsection{Universal Feature Interaction Learning}\label{sec:ufi}
The core of our proposed UFIN is to learn the universal feature interactions for intelligently acquiring the generalized collaborative knowledge across diverse domains.
To this end, our approach is to model the adaptive feature interactions based on the generated universal features to capture the common collaborative patterns.	
Furthermore, to promote feature interaction learning, we introduce a framework for distilling knowledge to guide the model's learning process and subsequently enhance its performance.	

\subsubsection{Adaptive Feature Interaction Learning at a Single Domain}
For learning universal feature interactions, an important issue is how to accurately model the interactions orders/forms within each domain,
as different domains typically correspond to varying feature relationships.
Traditional methods manually design a maximal order and further remove the useless interactions from them, \eg FM~\cite{rendle2010factorization} empirically enumerates all second-order feature interaction terms.
These approaches not only results in inaccurate modeling of the underlying true feature interactions in real-world scenarios but also limits model transferability.
As a promising approach, recent study EulerNet~\cite{tian2023eulernet} proposes to model the \emph{adaptive} feature interactions, \ie the interaction forms are automatically learned from data, allowing for arbitrary orders and a flexible number of terms.
Given the input features $\tilde{\bm E} = \{\tilde{\bm e}_1, ..., \tilde{\bm e}_{n_u}\}$ (See Eq.~\eqref{eq:unf}), the adaptive feature interaction learning function of EulerNet~\cite{tian2023eulernet} is formulated as:
\begin{equation}\label{eq-ex-int}
\mathcal{F}(\tilde{\bm E} ; \mathcal{A}) = \bm w^{\top}\sum_{\bm \alpha \in \mathcal{A}}  \tilde{\bm e}_1^{\alpha_1} \odot \tilde{\bm e}_2^{\alpha_2} \odot ... \odot \tilde{\bm e}_{n_u}^{\alpha_{n_u}},
\end{equation}
where $\bm \alpha = [\alpha_1, \alpha_2, ..., \alpha_{n_u}]$ denotes the learnable order parameter of each feature, $\mathcal{A}$ is the parameter set of all the learnable orders, and $\bm w$ is a transition vector for generating a scalar result.
For a given domain, the underlying true feature interaction forms are automatically learned from the parameter $\mathcal{A}$, \eg the interactions of FMs~\cite{rendle2010factorization} can be learned by $\mathcal{A} = \{ \bm{\alpha} | \sum_{j=1}^{m} \alpha_j = 2, \forall \alpha_j \in \{0, 1\}\}$.
Previous works~\cite{cheng2020adaptive, cai2021arm} face challenges in achieving this, because when the embedding $\bm e_j$ contains negative values, the order $\alpha_j$ must be set to an integer value to avoid invalid operation (\eg $(-1)^{0.5}$).
As a solution, EulerNet~\cite{tian2023eulernet} leverages Euler's formula to learn the feature interactions in a {complex vector space} that enables the efficient learning of arbitrary-order feature interactions, without additional restrictions (\eg \emph{non-negative} embedding or \emph{integer} order).

In our case, we utilize EulerNet~\cite{tian2023eulernet} to adaptively learn the underlying true feature interaction orders/forms within each given domain. We mainly transfer the knowledge of interaction orders (\ie $\mathcal{A}$) to enhance predictions in a target domain.

\subsubsection{Multi-Domain Feature Interaction Learning}\label{sec: fmoe}
In the above, we have discussed the feature interaction learning within a single domain.
 In this section, we aim to generalize the methodology for adapting the modeling of feature interactions across multiple domains. Intuitively, we can train distinct models independently to adapt to the distribution of each domain. However, this approach fails to grasp the interconnectedness between diverse domains, leading to challenges in the cold-start scenario. Our objective is to acquire knowledge of the universal feature interactions that encompass the shared collaborative patterns between different domains. Our proposed solution entails the introduction of multiple sets of interaction orders (\ie $\mathcal{A}$), each of which learns the underlying true feature interaction orders for a single domain, and shares some of them across different domains to acquire knowledge of the common collaborative patterns.

In practice, we employ an MoE model to implement our idea. 
Given $L$ domains, we introduce $L$ experts, each expert $j$ is implemented by EulerNet~\cite{tian2023eulernet} with the learnable order parameters $\mathcal{A}_j$.
All experts share the input embedding $\tilde{\bm E} = \{\tilde{\bm e}_1, ..., \tilde{\bm e}_{n_u}\}$ (See Eq.~\eqref{eq:unf}), combined by a gating router based on the semantic representation $\tilde{\bm z}$ (See Eq.~\eqref{eq:lat}):
\begin{align}\label{eq:fiexp}
    \zeta &= \sum_{j = 1}^L \mathcal{F}({\tilde{\bm E}} ; \mathcal{A}_j) \cdot \tilde{\bm g}_j,\\
    \tilde{\bm g} &= \mathrm{TopK}\Big(\mathrm{Softmax}(\tilde{\bm W}_g \tilde{\bm z})\Big),\label{eq:esele}
\end{align}
where $\mathrm{TopK}(\cdot)$ retains only the top-K elements of the input tensor and sets all other elements as zero, $\mathcal{A}_j$ (\emph{learnable parameter}) is the interaction order of $j$-th expert, 
$\mathcal{F}(\cdot)$ is the feature interaction function (See Eq.~\eqref{eq-ex-int}) of EulerNet~\cite{tian2023eulernet}, $\tilde{\bm g}_j$ is the $j$-th combination weight, $\tilde{\bm W}_g \in \mathbb{R}^{L \times d_V}$ is the router weight, and $\zeta$ is the output logits.
For an instance in a given domain $u$, we use its corresponding semantic representations $\tilde{\bm z}$ to select $K$ experts with the order sets $\mathcal{S}_u = \{\mathcal{A}^u_j\}_{j=1}^K$ for collaboratively learning the feature interactions.
Formally we set $K > \lceil{L/2}\rceil$, and we have the following finding:

\begin{theorem}\label{thee}
	Given $L$ domains $\mathcal{D} = \{D_1, D_2, ..., D_L\}$ and $L$ experts $\mathcal{A} = \{\mathcal{A}_1, \mathcal{A}_2, ..., \mathcal{A}_L\}$, each domain $D_u$ select $K$ experts $\mathcal{S}_u = \{\mathcal{A}_1^u, \mathcal{A}_2^u, ..., \mathcal{A}_K^u\}$ from $\mathcal{A}$, \ie $\mathcal{S}_u \subseteq \mathcal{A}$. If $K > \lceil{L/2}\rceil$, then for any given domain $D_u$ and $D_v$, the set of the selected experts $\mathcal{S}_u$ and $\mathcal{S}_v$ must have a same element, \ie $\mathcal{S}_u \cap \mathcal{S}_v \neq \emptyset, \forall u \neq v$.
 See proofs in Section~\ref{sec:pro}.
\end{theorem}

It demonstrates that for any given domains $u$ and $v$, there exist at least one shared expert that learns the common feature interactions.
Therefore, our approach can capture the common feature interactions between arbitrary domain pairs, thereby capable of learning the generalized feature relationship across all domains. 
Finally, we apply the sigmoid function on the logits (See Eq.~\eqref{eq:fiexp}) to obtain the prediction:
\begin{align}\label{eq:ot}
    \hat{y} = \textrm{Sigmoid}(\zeta).
\end{align}

As mentioned in the work~\cite{sheng2021one}, a good multi-domain CTR model should contain the features that depict the domain-specific information.
For this purpose, we can further incorporate a \textbf{\emph{feature adaptor}}\label{sec:fad}, which takes as input  the domain-specific features, to precisely capture the distinct characteristics of each domain.
Following existing multi-domain methods~\cite{sheng2021one}, we add the output logits of the feature adaptor (\ie $\zeta_{f}$) to our approach (\ie $\zeta$) for prediction:
\begin{align}\label{eq:ott}
    \hat{y} = \textrm{Sigmoid}(\zeta + \zeta_f).
\end{align}

It allows the flexibility to choose any method.
Here we employ the simplest regression model (\eg LR~\cite{richardson2007predicting}) as the feature adaptor to  ensure the efficiency of the model.

\subsubsection{Knowledge Distillation Enhanced Training}\label{sec:kd}
With the above approaches, UFIN is able to learn the universal feature interactions based on the representations generated by LLMs.
However, as existing study shows~\cite{li2023ctrl}, it is challenging for the representations of LLMs to capture the feature co-occurrence correlation that results in the poor performance compared with traditional collaborative models.
To facilitate our approach in capturing the underlying collaborative patterns, we propose a multi-domain knowledge distillation framework that promotes feature interaction learning.

\paratitle{Multi-Domain Guided Distillation.}
In the knowledge distillation framework, for each domain, we pre-train a feature interaction model as the \emph{guided network} to learn the domain-specific collaborative patterns.
As such, multiple guided networks from different domains are incorporated as {teacher model} to supervise the training of our approach.
It allows flexibility to choose any teacher model for each domain, and we specifically employ  the EulerNet~\cite{tian2023eulernet} as the teacher with consideration for the consistency. 
Following existing studies~\cite{zhu2020ensembled, tian2023directed}, we use MSE loss to align the output logits between the guided network and our approach:
\begin{align}
    \mathcal{L}_{KD} = \sum_{p=1}^M\sum_{i = 1}^{N_p} ||\zeta^G_{p,i} - \zeta_{p,i}||^2,
\end{align}
where $\zeta^G_{p,i}$ and $\zeta_{p,i}$ (see Eq.~\eqref{eq:fiexp}) are the logits of the guided network and our proposed UFIN for the $i$-th instance in the $p$-th domain.
Note that the guided networks are only used for auxiliary training, which are discard during inference, thus ensuring the transferability of our approach.

\paratitle{Multi-Task Learning.}
To further promote the utilization of LLMs in the CTR prediction task, 
 we incorporate the commonly adopted binary cross-entropy loss:
\begin{align}
    \mathcal{L}_{CTR} = \sum_{p=1}^M\sum_{i=1}^{N_p}\bigg(y_i^p\log(\hat{y}_i^p)+(1-y_i^p)\log(1-\hat{y}_i^p)\bigg),
\end{align}
where $y_i^p$ and $\hat{y}_i^p$ are the ground-truth label and predicted result of $i$-th instance in the $p$-th domain.
As for the model training, we adopt a multi-task training strategy to jointly optimize the knowledge distillation loss and the binary classification loss:
\begin{align}
    \mathcal{L} = \mathcal{L}_{KD} + \mathcal{L}_{CTR}.
\end{align}

\subsection{Discussion} \label{sec:approach}
In the literature, a number of CTR models have been proposed.
To better highlight the novelty and difference of our approach, we make a brief comparison of different CTR methods.
For ID-based methods, such as STAR~\cite{sheng2021one}, they rely on the ID features to develop the prediction, which impairs the inherent semantic of features that makes them incapable of being applied in the new platforms.
While for semantic methods, such as P5~\cite{geng2022recommendation}, attempt to leverage the world knowledge of LLMs to uplift the prediction.
The primary challenge of these approaches lies in their inability to capture collaborative signals, resulting in poor performance.	
Although CTRL~\cite{li2023ctrl} proposes a contrastive learning framework to transfer the knowledge of LLMs to a domain-specific collaborative model.
Due to the limited capacity of the collaborative model, it cannot adequately learn the common knowledge across different domains and suffers the seesaw phenomenon (See Section~\ref{sec: overall}).
In contrast, our model is more \emph{universal} in the application of multi-domain or cross-platform settings, naturally integrating the semantic knowledge of LLMs and collaborative knowledge of interactions.
The overall comparison is presented  in Table~\ref{tab:cmp}.

\begin{table}[h]
\centering
\small
\caption{Comparison of different CTR methods.}
\label{tab:cmp}
\resizebox{1.\columnwidth}{!}{
\begin{tabular}{@{}lcccc@{}}
\toprule
\multirow{2}{*}{Methods} & \multicolumn{2}{c}{Transfer Learning} & \multicolumn{2}{c}{Knowledge Patterns}
\\ \cmidrule(l){2-3} \cmidrule(l){4-5}
     & Multi-Domain & Cross-Platform & Collaborative & Semantic \\ \midrule
STAR~\cite{sheng2021one} & \textcolor{teal}{\CheckmarkBold} & \textcolor{purple}{\XSolidBrush} & \textcolor{teal}{\CheckmarkBold} & \textcolor{purple}{\XSolidBrush}\\ 
CTRL~\cite{li2023ctrl}    & \textcolor{purple}{\XSolidBrush} & \textcolor{purple}{\XSolidBrush} & \textcolor{teal}{\CheckmarkBold} & \textcolor{teal}{\CheckmarkBold}\\ 
P5~\cite{geng2022recommendation}   & \textcolor{teal}{\CheckmarkBold} &\textcolor{teal}{\CheckmarkBold} & \textcolor{purple}{\XSolidBrush} & \textcolor{teal}{\CheckmarkBold}\\
UFIN (ours) & \textcolor{teal}{\CheckmarkBold}  & \textcolor{teal}{\CheckmarkBold} & \textcolor{teal}{\CheckmarkBold} & \textcolor{teal}{\CheckmarkBold}\\ \bottomrule
\end{tabular}
}
\end{table}


\section{EXPERIMENTS}
In this section, we first introduce the experimental settings and then give the results and analysis.

\subsection{Experimental Settings}
We present the experimental settings, including the datasets,  baseline approaches, the details of hyper-parameters, and evaluation metrics.

\subsubsection{Datasets}
It is worth mentioning that the majority of alternative public datasets cannot be utilized to evaluate our model due to their inclusion of solely anonymous features and the absence of semantic features.
To evaluate the performance of our model, we conduct experiments on the Amazon\footnote{https://nijianmo.github.io/amazon/index.html} and MovieLens-1M\footnote{https://grouplens.org/datasets/movielens} datasets in both multi-domain setting and cross-platform setting.
The statistics of datasets are summerized in Table~\ref{tab:datasets}.

\begin{itemize}
    \item 
\textbf{Amazon}~\cite{ni2019justifying}  is a widely-used dataset for recommender systems research. We keep the five-core datasets and select seven subsets for the multi-domain setting (\ie \emph{"Movies and TV"}, \emph{"Books"}, \emph{"Electronics"}, \emph{"Office Products"}, \emph{"Musical Instruments"}, \emph{"Toys and Games"} and \emph{"Grocery and Gourmet Food"}).

    \item 
 \textbf{MovieLens-1M} is a movie recommendation dataset, which does not contain overlapped users or items with Amazon. We use this dataset to evaluate the performance in a cross-platform setting.
\end{itemize}

  \begin{table}[!h]
    \centering
    \caption{The statistics of datasets.} 
    \label{tab:datasets}
    \begin{tabular}{l|rrr}
      \toprule
      \textbf{Dataset}& $\#$ \textbf{Users} & $\#$ \textbf{Items} & $\#$ \textbf{Interactions} \\
      \hline \hline
      \textbf{Amazon} & 1,002,827 & 2,530,874 & 7,427,505 \\
      -~Movies & 295,908 & 40,792 & 2,141,592 \\
      -~Books & 585,167 & 191,826 & 4,077,731\\
      -~Electronics & 184,876 & 26,336 & 780,698 \\
      -~Food & 14,552 & 5,474 & 86,518\\
      -~Instruments & 1,429 & 736 & 7,835\\
      -~Office & 4,890 & 1,805 & 36,572\\
      -~Toys & 19,231 & 9,142 & 116,559\\
      \hline
      \textbf{MovieLens-1M} & 6,041 & 3,669 & 739,012 \\
    \bottomrule
  \end{tabular}
  \end{table}

\subsubsection{Compared Models}
We compare the UFIN with twelve state-of-the-art methods, including single-domain methods, multi-domain methods and semantic methods.

\noindent \textbf{Single-Domain Methods:}

\begin{itemize}
    \item  DeepFM~\cite{guo2017deepfm} combines traditional second-order factorization machines with a feed-forward neural network (MLP), utilizing a shared embedding layer.

\item DCNV2~\cite{wang2021dcn} is the state-of-the-art model which   captures the high-order feature interactions by taking kernel product operation of concatenated feature vectors.

\item xDeepFM~\cite{lian2018xdeepfm} introduces the Compressed Interaction Network (CIN), which forms the cornerstone of the xDeepFM model. It employs the outer product of stacked feature matrices at the vector-wise level. Additionally, a feed-forward neural network is integrated to enhance the feature representations.

\item EulerNet~\cite{tian2023eulernet} utilize the Euler's formula to map the feature representations from the real vector space in a \emph{complex vector space}, which provides a way to cast the feature interaction orders to the linear coefficients.
As such, it can automatically learn the arbitrary-order feature interactions from data.
\end{itemize}

\noindent \textbf{Multi-Domain Methods:}

\begin{itemize}
\item SharedBottom~\cite{ruder2017overview} is a multi-task model that parameterizes the bottom layers. In our implementation, we apply SharedBottom to share the embedding layer, while each domain retains its distinct fully-connected network that remains unshared.

\item MMoE~\cite{ma2018modeling} implicitly captures task relationships in multi-task learning, accommodating diverse label spaces for different tasks. We extend MMoE to the context of multi-domain CTR prediction, configuring each expert as a fully-connected network. The count of experts matches the domain count, and MMoE additionally incorporates gating networks for each domain. These networks process input features and yield softmax gates, dynamically combining experts with varying weights.

\item HMOE~\cite{tang2020progressive} extends MMoE to scenario-aware experts using a gradient cutting trick to explicitly encode scenario correlations.

\item AESM$^2$~\cite{zou2022automatic} introduces an innovative expert selection algorithm that automatically identifies scenario-/task-specific and shared experts for each input.

\item STAR~\cite{sheng2021one} proposes a star topology architecture, which consists of a shared center network and multiple domain-specific networks for the adaptation of multi-domain distributions.

\item PEPNet~\cite{chang2023pepnet} introduces an embedding personalized network to align multi-domain feature representations.
The common knowledge from different domains can be learned in a shared Gated-Net.
\end{itemize}

\noindent \textbf{Semantic Methods:} 
\begin{itemize}
\item P5~\cite{geng2022recommendation}
is a semantic-based recommendation
model that transforms diverse recommendation tasks into text generation
tasks through prompt learning, leveraging T5~\cite{raffel2020exploring} as the underlying model.

\item CTRL~\cite{li2023ctrl} proposes to combine the collaborative signals and semantic knowledge in a contrastive learning framework.
Specifically, it utilizes the knowledge of LLMs to enhance the learning of a collaborative. Only the collaborative model will online.
\end{itemize}

For our approach, since there are no effective user-side textual features (\eg only \emph{user\_id} available on the Amazon dataset), we incorporate the \emph{user\_id} features into our approach (See Section~\ref{sec: faf}).
We introduce two versions of our approach: (1) \textbf{\underline{UFIN$_{t}$}} denotes the model using only item text; (2) \textbf{\underline{UFIN$_{t+f}$}} denotes the model using  item text and integrating a feature adaptor (See Section~\ref{sec:fad}).

\subsubsection{Implementation Details} \label{sec: impd}
The dataset is randomly split into 80\% for training, 10\% for validation and 10\% for testing.
Following the work~\cite{wang2021dcn}, we convert the ratings of 4-5 as label 1; ratings of 1-2 as label 0; and remove the ratings of 3.
Following the work~\cite{chang2023pepnet}, we separately train each single-domain model in each domain to report their best results.
For multi-domain evaluation, we mix all train data of the Amazon dataset to train the multi-domain models and semantic models, and  evaluate their performance in each given domain.
For cross-platform evaluation, we first pre-train each semantic model on the mixed Amazon dataset, and then fine-tune them on the MovieLens-1M dataset.

The baseline implementation follows FuxiCTR~\cite{10.1145/3459637.3482486}.
For each method, the grid search is applied to find the optimal settings. 
The embedding size is 16, learning rate is tuned in \{1e-3, 1e-4, 1e-5\}, and batch\_size is 1024. 
The optimizer is Adam~\cite{kingma2014adam}.
The $L_2$ penalty weight is in \{1e-3, 1e-5, 1e-7\}. 
The default settings for MLP components are:
(1) the architecture of hidden layer is $400 \times 400 \times 400$; 
(2) the dropout rate is 0.2.
For DeepFM, the hidden layer of MLP is in \{$128 \times 128 \times 128$, $256 \times 256 \times 256$, $400 \times 400 \times 400$\}.
For xDeepFM, the depth of CIN is in \{1, 2, 3, 4, 5\} and the hidden size is in \{100, 200, 400\}.
For DCNV2, the depth of CrossNet is in \{1, 2, 3, 4, 5\}.
For EulerNet, the number of Euler interaction layer is in \{1, 2, 3, 4, 5\}, and the number of order vectors is in \{5, 10, 20, 30, 40\}.
For SharedBottom, the hidden layer of MLP for each domain is in \{$128 \times 128 \times 128$, $256 \times 256 \times 256$, $400 \times 400 \times 400$\}.
For MMoE, the numer of experts is in $\{2,4,8,10\}$, the hidden layer of gating network is in \{$64 \times 64 \times 64$, $128 \times 128 \times 128$, $400 \times 400 \times 400$\}.
For HMoE, the numer of experts is in $\{2,4,8,10\}$, the hidden layer of scenario network is in \{$64 \times 64 \times 64$, $128 \times 128 \times 128$, $400 \times 400 \times 400$\}.
For AMSE$^2$, the numer of experts is in $\{2,4,8,10\}$, the layer number is in $\{1,2,3,4,5\}$.
For STAR, the hidden layer of the auxiliary networks and topology network is in \{$128 \times 128 \times 128$, $256 \times 256 \times 256$, $400 \times 400 \times 400$\}.
For PEPNet, the gate hideen dim is in $\{16, 32, 48, 64, 80\}$, and hidden units of PPNet is in \{$128 \times 128 \times 128$, $256 \times 256 \times 256$, $400 \times 400 \times 400$\}.
For P5, we use T5 as the base model.
For CTRL, we try to use RoBERTa and T5 as the semantic model. The collaborative model is AutoInt, the depth, number of head and attention size is 2, 2, 40 respectively.
For our approach, the number of experts in the semantic fusion MoE and feature interaction MoE is equal to the number of domains (\ie $L = 7$).
The universal feature fields $n_u$ is set to $7$ and dimension $d$ is set to 16.
The $K$ value of $TopK$ function is 5.
The SLP hidden size is $128$.
For the EulerNet expert, the layer number is 1, and the number of order vectors is 7.

\subsubsection{Evaluation Metrics}
We use two popular metrics to evaluate the model performance: AUC and LogLoss.

$\bullet$ AUC~\cite{lobo2008auc} (Area Under the Curve) is a metric used to evaluate the performance of binary classification or ranking models, particularly in tasks where the prediction involves ranking items. It quantifies the ability of a model to correctly rank higher the positive instances compared to negative instances. A higher AUC value indicates better model performance.

In the context of ranking tasks, AUC is calculated using the following formula:

\[ \text{AUC} = \frac{\sum_{(i,j) \in S} \mathbbm{1}(s_i > s_j)}{|S|} \]

Where \( S \) is the set of all pairs of instances \( (i, j) \) where \( i \) is a positive instance and \( j \) is a negative instance, \( s_i \) and \( s_j \) represent the predicted scores or ranks of instances \( i \) and \( j \), respectively. \( \mathbbm{1}(s_i > s_j) \) is the indicator function that equals 1 if the predicted score of \( i \) is greater than that of \( j \), and 0 otherwise. \( |S| \) denotes the total number of pairs in the set \( S \).
In this context, AUC measures the probability that a randomly chosen positive instance will be ranked higher than a randomly chosen negative instance according to the model's predictions.

$\bullet$  LogLoss~\cite{buja2005loss}, also known as logarithmic loss or cross-entropy loss, is a common evaluation metric used in binary classification and ranking tasks. It measures the difference between predicted probabilities and actual binary labels. In the context of ranking tasks, LogLoss is often used to assess the quality of predicted relevance scores for different items.
In a ranking task, the LogLoss formula can be expressed as:

\[ \text{LogLoss} = -\frac{1}{N} \sum_{i=1}^{N} \Big( y_i \cdot \log(p_i) + (1 - y_i) \cdot \log(1 - p_i) \Big) \]

where \(N\) is the number of instances (samples), \(y_i\) is the true binary label (0 or 1) for the \(i\)-th instance, and \(p_i\) is the predicted probability of the positive class for the \(i\)-th instance.
LogLoss penalizes larger differences between predicted probabilities and true labels. 
It encourages the model to make confident predictions that are close to the true labels. 
A lower LogLoss value indicates better alignment between predicted probabilities and actual labels, reflecting better model performance.

\subsection{Overall Performance}\label{sec: overall}

\begin{table*}[!ht]
\centering

\caption{Performance comparison of different CTR models.
``Improv.'' indicates the relative improvement ratios of our proposed approach over the best baseline.
``*'' denotes that statistical significance for $p < 0.01$ compare to the best baseline. \textbf{Note that a higher AUC or lower Logloss at 0.001-level is regarded significant}, as stated in previous studies~\cite{song2019autoint, cheng2016wide, guo2017deepfm}. }
\label{tab:exp-main}
\resizebox{2.05\columnwidth}{!}{
\begin{tabular}{@{}ccccccccccccccccccr@{}}
\toprule
\multirow{2}{*}{Evaluation} & \multirow{2}{*}{Dataset} & \multirow{2}{*}{Metric} & \multicolumn{4}{c}{Single-Domain Methods} & \multicolumn{6}{c}{Multi-Domain Methods} & \multicolumn{4}{c}{Semantic Methods} & \multirow{2}{*}{ Improv.} \\
\cmidrule(l){4-7} \cmidrule(l){8-13} \cmidrule(l){14-17}
& & & DeepFM & xDeepFM  & DCNV2 & EulerNet & SharedBottom & MMoE & HMoE & AESM$^2$ & STAR & PEPNet & P5 & CTRL & UFIN$_t$ & UFIN$_{t+f}$ & \\ \midrule \midrule
\multirow{14}{*}{\shortstack{Multi-\\Domain}} &
\multirow{2}{*}{Movies} &
     AUC & \underline{0.8568} & 0.8565 & 0.8537 & 0.8550 & 0.8465 & 0.8440 & 0.8445 & 0.8477   & 0.8448 & 0.8456 & 0.6440 & 0.8435 & 0.8515 & \textbf{0.8582$^*$} &  $+0.16\%$ \\
 & & LogLoss & \underline{0.2670} & 0.2738 & 0.2698 & 0.2685 & 0.2752 & 0.2751 & 0.2755 & 0.2726 & 0.2738 & 0.2746 & 0.3760 & 0.2736 & 0.2741 & \textbf{0.2649$^*$} &  $+0.79\%$\\
\cmidrule(l){2-18}
 & \multirow{2}{*}{Books} &
     AUC & 0.8107 & 0.8094 & 0.8118 & \underline{0.8166} & 0.8044 & 0.8095 & 0.8097 & 0.8108 & 0.8057 & 0.8080 & 0.5740 & 0.8058 & 0.8130 & \textbf{0.8215$^*$} & $+0.60\%$ \\
 & & LogLoss & 0.2486 & 0.2492 & 0.2484 & \underline{0.2458} & 0.2528 & 0.2509 & 0.2500 & 0.2489 & 0.2507 & 0.2505 & 0.3332 & 0.2549 & 0.2514 & \textbf{0.2437$^*$} & $+0.85\%$\\
 \cmidrule(l){2-18}
 & \multirow{2}{*}{Electronics} &
     AUC & 0.7493 & 0.7504 & 0.7508 & \underline{0.7524} & 0.7437 & 0.7455 & 0.7420 & 0.7438 & 0.7277 & 0.7386 & 0.5762 & 0.7376 & 0.7337 & \textbf{0.7532} &  $+0.11\%$\\
 & & LogLoss & 0.3207 & 0.3192 & 0.3186 & \textbf{0.3181} & 0.3318 & 0.3217 & 0.3227 & 0.3218 & 0.3272 & 0.3272 & 0.3751 & 0.3218 & 0.3336 & \underline{0.3185} &  $-$\\
 \cmidrule(l){2-18}
 & \multirow{2}{*}{Food} &
     AUC & 0.7198 & 0.7208 & 0.7204 & 0.7188 & 0.6833 & 0.7103 & 0.7117 & 0.7154 & 0.7139 & 0.7135 & 0.5899 & 0.7188 & \underline{0.7231} & \textbf{0.7322$^*$} & $+1.58\%$\\
 & & LogLoss & 0.2938 & 0.2902 & 0.2944 & 0.3064 & 0.2891 & 0.2841 & 0.2829 & 0.2822 & \underline{0.2810} & 0.2836 & 0.3658 & 0.3034 & 0.2903 & \textbf{0.2774$^*$} &  $+1.28\%$\\
 \cmidrule(l){2-18}
 & \multirow{2}{*}{Instruments} &
     AUC & 0.5811 & 0.5879 & 0.5789 & 0.6056 & 0.6137 & 0.6275 & 0.6301 & 0.6587 & 0.6618 & \textbf{0.6780} & 0.5623 & 0.5978 & 0.6635 & \underline{0.6768} &  $-$\\
 & & LogLoss & 0.4921 & 0.4751 & 0.4514 & 0.3070 & 0.2115 & 0.1809 & 0.1717 & \underline{0.1714} & \textbf{0.1711} & 0.1722 & 0.2494 & 0.4929 & {0.1720} & 0.1798 &  $-$\\

 \cmidrule(l){2-18}
 & \multirow{2}{*}{Office} &
     AUC & 0.7391 & 0.7405 & 0.7425 & 0.7401 & 0.7187 & 0.7334 & 0.7524 & 0.7522  & \underline{0.7788} & 0.7554 & 0.5755 & 0.7432 & 0.7590 & \textbf{0.7812$^*$} &  $+0.31\%$\\
 & & LogLoss & 0.2334 & 0.2503 & 0.2077 & 0.2353 & 0.2116 & 0.2038 & 0.2018 & 0.2050  & \underline{0.1974} & 0.2024 & 0.2743 & 0.2721 & 0.2179 & \textbf{0.1945$^*$} & $+1.47\%$\\
 \cmidrule(l){2-18}
 & \multirow{2}{*}{Toys} &
     AUC & 0.7781 & {0.7821} & \underline{0.7856} & 0.7820 & 0.7620 & 0.7666 & 0.7799 & 0.7806 & 0.7725 & 0.7717 & 0.5772 & 0.7623 & 0.7768 & \textbf{0.7964$^*$} & $+1.37\%$\\
 & & LogLoss & 0.2196 & {0.2177} & 0.2207 & 0.2277 & 0.2554  & 0.2164 & 0.2141 & \underline{0.2116} & 0.2124 & 0.2145 & 0.2937 & 0.2218 & 0.2262 & \textbf{0.2061$^*$} & $+2.59\%$\\
 \midrule
\multirow{2}{*}{\shortstack{Cross-\\Platform}} &
\multirow{2}{*}{\shortstack{MovieLens-1M}} &
     AUC & 0.8973 & 0.8969 & 0.8989 & 0.9018 & 0.8969 & 0.8967 & 0.8978 & 0.8963 & 0.8964 & 0.8970 & 0.7840 & 0.8970 & \underline{0.9024} & \textbf{0.9029$^*$} & $+0.12\%$\\
 & & LogLoss & 0.3166 & 0.3189 & 0.3147 & \underline{0.3086} & 0.3165 & 0.3183 & 0.3183 & 0.3206 & 0.3167 & 0.3151 & 0.4482 & 0.3155 & 0.3096 & \textbf{0.3053$^*$} &  $+1.07\%$\\
\bottomrule
\toprule
\multirow{2}{*}{\shortstack{Efficiency}} &
\multirow{1}{*}{\shortstack{Amazon}} & Latency (ms) & 0.0238 & 0.0345 & 0.0270 & 0.0323 & 0.0333 & 0.0357 & 0.1111 & 0.1250 & 0.0400 & 0.0370 & 6.5737 & 0.0714 & 0.0385 & 0.0476 & $-$ \\
&\multirow{1}{*}{\shortstack{MovieLens-1M}} & Latency (ms) & 0.0247 & 0.0352 & 0.0264 & 0.0349 & 0.0307 & 0.0394 & 0.1289 & 0.1470 & 0.0564 & 0.0477 & 6.5737 & 0.0852 & 0.0453 & 0.0594 & $-$ \\

 \bottomrule
\end{tabular}
}
\end{table*}

We compare the proposed approach with twelve baseline methods on seven multi-domain datasets and one cross-platform dataset.
The overall performance is presented in Table~\ref{tab:exp-main}, and we have the following observations:

(1) Single-domain methods (\ie DeepFM~\cite{guo2017deepfm}, xDeepFM~\cite{lian2018xdeepfm}, DCNV2~\cite{wang2021dcn} and EulerNet~\cite{tian2023eulernet}) perform well on the Movies and Books datasets, but exhibit subpar performance
 on sparse domains with less interactions (\ie Instruments and Office).
This is consistent with the observations reported in PEPNet~\cite{chang2023pepnet}, where  single-domain methods exhibit varying performance across different domains.

(2) Multi-domain methods (\ie SharedBottom~\cite{ruder2017overview}, MMoE~\cite{ma2018modeling}, HMoE~\cite{tang2020progressive}, AESM$^2$~\cite{zou2022automatic}, STAR~\cite{sheng2021one} and PEPNet~\cite{chang2023pepnet}) achieve comparable performance on the Instruments and Office datasets, showing the effectiveness of multi-domain information sharing.

(3) For semantic methods, P5~\cite{geng2022recommendation} performs poorly across all datasets due to the limitation of LLMs in capturing collaborative signals. 
Besides, CTRL~\cite{li2023ctrl} achieves better performance by aligning the representations between LLMs and collaborative models, but it becomes less effective on the Instruments dataset, showing its limitation of capturing the relatedness between different domains.


(4) Our proposed approach achieves the best performance in almost all the cases.
It signifies that our suggested universal feature interactions are better suited for the adaptation of multi-domain distributions.
Further, the cross-platform evaluation results show that our approach can be effectively transferred to a new platform.
Besides, UFIN$_{t+f}$ has a great improvement over UFIN$_{t}$, showing the effectiveness of incorporating feature adaptor to learn domain-specific collaborative patterns.

Regarding the efficiency of the model, it can be observed that the latency of single-domain models (\ie DeepFM~\cite{guo2017deepfm}, xDeepFM~\cite{lian2018xdeepfm}, DCNV2~\cite{wang2021dcn} and EulerNet~\cite{tian2023eulernet}) is relatively small due to their simplistic architecture. 
In the case of SharedBottom~\cite{ruder2017overview}, MMoE~\cite{ma2018modeling}, STAR~\cite{sheng2021one}, and PEPNet~\cite{chang2023pepnet}, they are more efficient as their learning strategies are simpler. Conversely, the latency of AESM$^2$~\cite{zou2022automatic} and HMoE~\cite{tang2020progressive} is relatively large due to their intricate architecture and learning algorithm.
For semantic models, the latency of P5~\cite{geng2022recommendation} is much larger since the training process of PLMs is extremely time-consuming.
In contrast, CTRL~\cite{li2023ctrl} and our approach are much more efficient.
This is because we can cache the textual representations of the LLMs, and only the lightweight feature interaction backbone needs to be deployed in the inference stage, thereby preserving the efficient online inference akin to traditional recommendation models.

\subsection{Further Analysis}

\subsubsection{Zero-shot Learning Analysis}
\begin{figure}[!h]
  \centering
  \includegraphics[width=1.\linewidth]{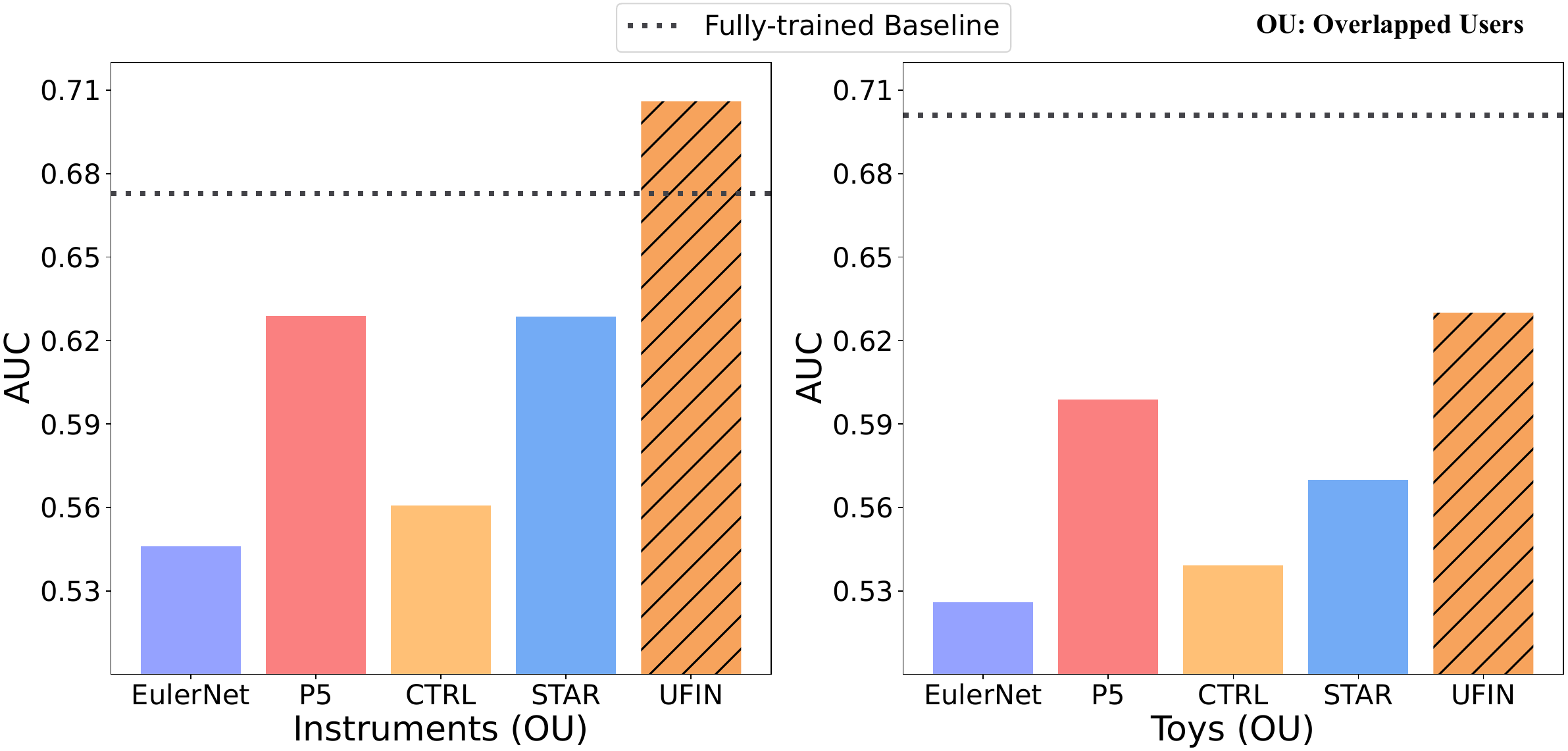}
  
  \caption{Performance comparison under the zero-shot setting.}
  \label{fig:zero-shot}
\end{figure}

To show the transfer learning ability of our approach, we evaluate the zero-shot performance of five methods (\ie EulerNet~\cite{tian2023eulernet}, P5~\cite{geng2022recommendation}, CTRL~\cite{li2023ctrl}, STAR~\cite{sheng2021one} and our proposed UFIN), and compare the results to the best performance of fully-trained single-domain methods.
In this setting, we train the model on three pre-trained datasets (\ie \emph{Movies}, \emph{Books} and \emph{Electronics}) and directly test them on two downstream datasets (\ie \emph{Instruments} and \emph{Toys}) without further training.
The downstream datasets retain only the interactions involving  \textbf{overlapped users} from the pre-trained datasets.
As shown in Figure~\ref{fig:zero-shot}, UFIN achieves the best zero-shot performance on both datasets.
On the Instruments dataset, UFIN performs even better than the fully-trained single-doamin methods.
These results demonstrate the strong transferability and inductive capability of our approach in learning general knowledge across different domains.

\subsubsection{Cross-Platform Learning Analysis}\label{sec:cp}
In this part, we investigate the effectiveness of the proposed LLM based representation approach.
Specially, we examine whether the model pre-trained on the Amazon datasets performs better than those without pre-training.
We compare the performance of the following representation methods:  (1) \underline{Embedding Look-up}: use an embedding vector to represent each feature, (2) \underline{PLM based encoding}:  encode the textual data with PLMs (\ie BERT~\cite{devlin2018bert}, T5~\cite{raffel2020exploring} and FLAN-T5~\cite{chung2022scaling}).

Experimental results are shown in Table~\ref{tab:trans}. 
We observe that PLM-based methods perform better when pre-training is applied, whereas the Embedding Look-up method is negatively affected by pre-training.
It indicates that natural language is more promising as the general representations across different scenarios.
Furthermore, the approach with T5~\cite{raffel2020exploring} and FLAN-T5~\cite{chung2022scaling} largely outperforms other methods, showing the excellent language modeling capacity of LLMs.
\begin{table}[!h]
  \centering
  \caption{Cross-platform performance on the MovieLens-1M dataset.}
  \label{tab:trans}
  \resizebox{1.\columnwidth}{!}{
  \begin{tabular}{c|cc|cc|c}
    \toprule
    \multirow{2}{*}{\textbf{Representation Methods}}&\multicolumn{2}{c|}{\textbf{w/o. Pre-train}}&\multicolumn{2}{c|}{\textbf{w. Pre-train}} & \multirow{2}{*}{\textbf{Effect.}}\\
    & AUC  & Log Loss & AUC & Log Loss \\
    \hline \hline
    Embedding Look-up & 0.8997 & 0.3102 & 0.8990 & 0.3109 & $-0.08\%$\\ 
    BERT~\cite{devlin2018bert} & 0.9001 & 0.3089 & 0.9009 & 0.3084 & $+0.09\%$\\
    T5~\cite{raffel2020exploring} & 0.9008 & 0.3105 & 0.9027 & 0.3054 & $+0.21\%$\\
    FLAN-T5~\cite{chung2022scaling}) & 0.9013 & 0.3074 & 0.9030 & 0.3050 & $+0.19\%$\\
  \bottomrule
\end{tabular}}
\end{table}

\subsubsection{Ablation Study}\label{exp: alb}
\begin{figure}[!h]
  \centering
  \subcaptionbox{Components}{
    \includegraphics[width=0.46\linewidth]{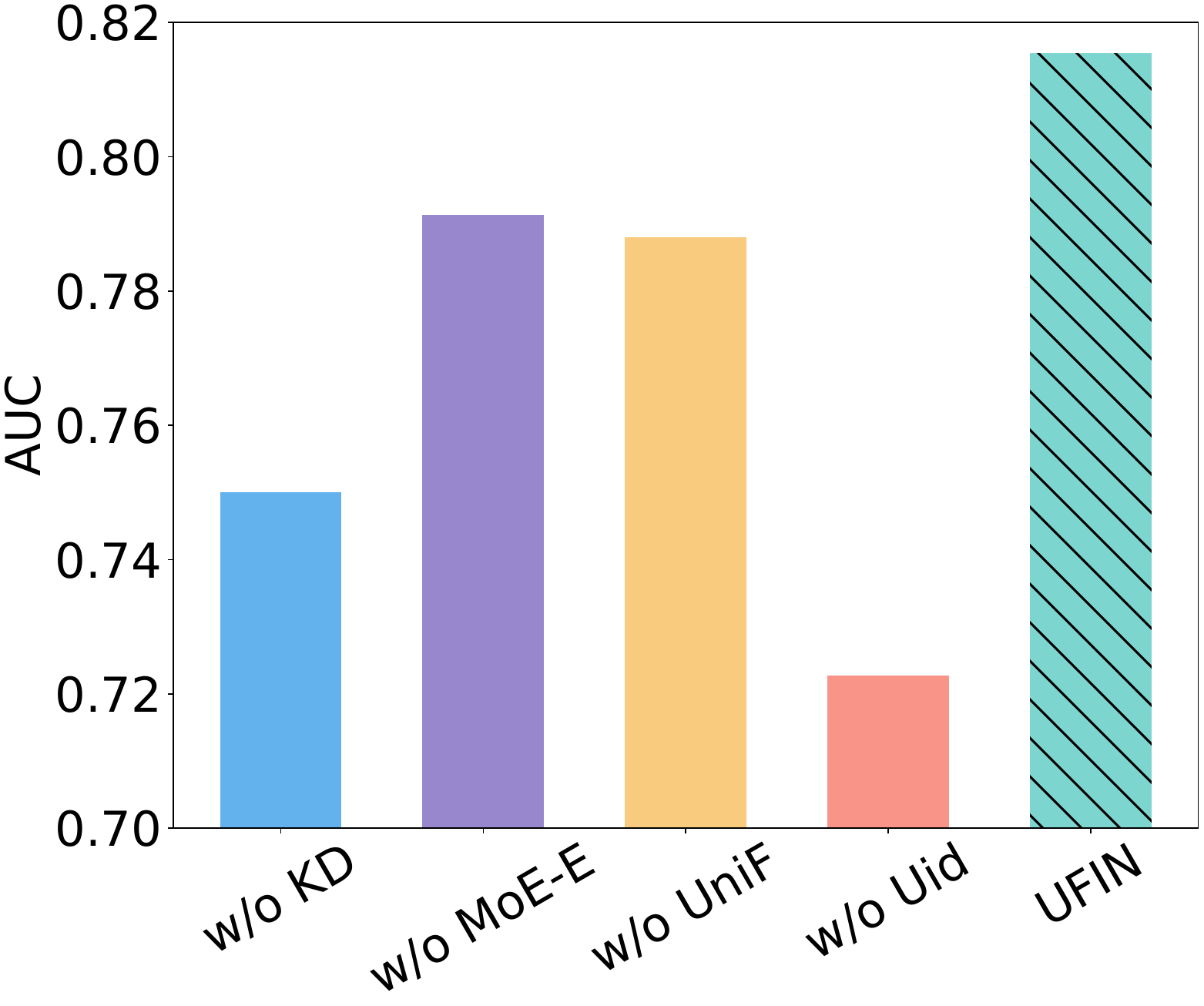}
  }
  \subcaptionbox{Feature Interaction Expert}{
    \includegraphics[width=0.46\linewidth]{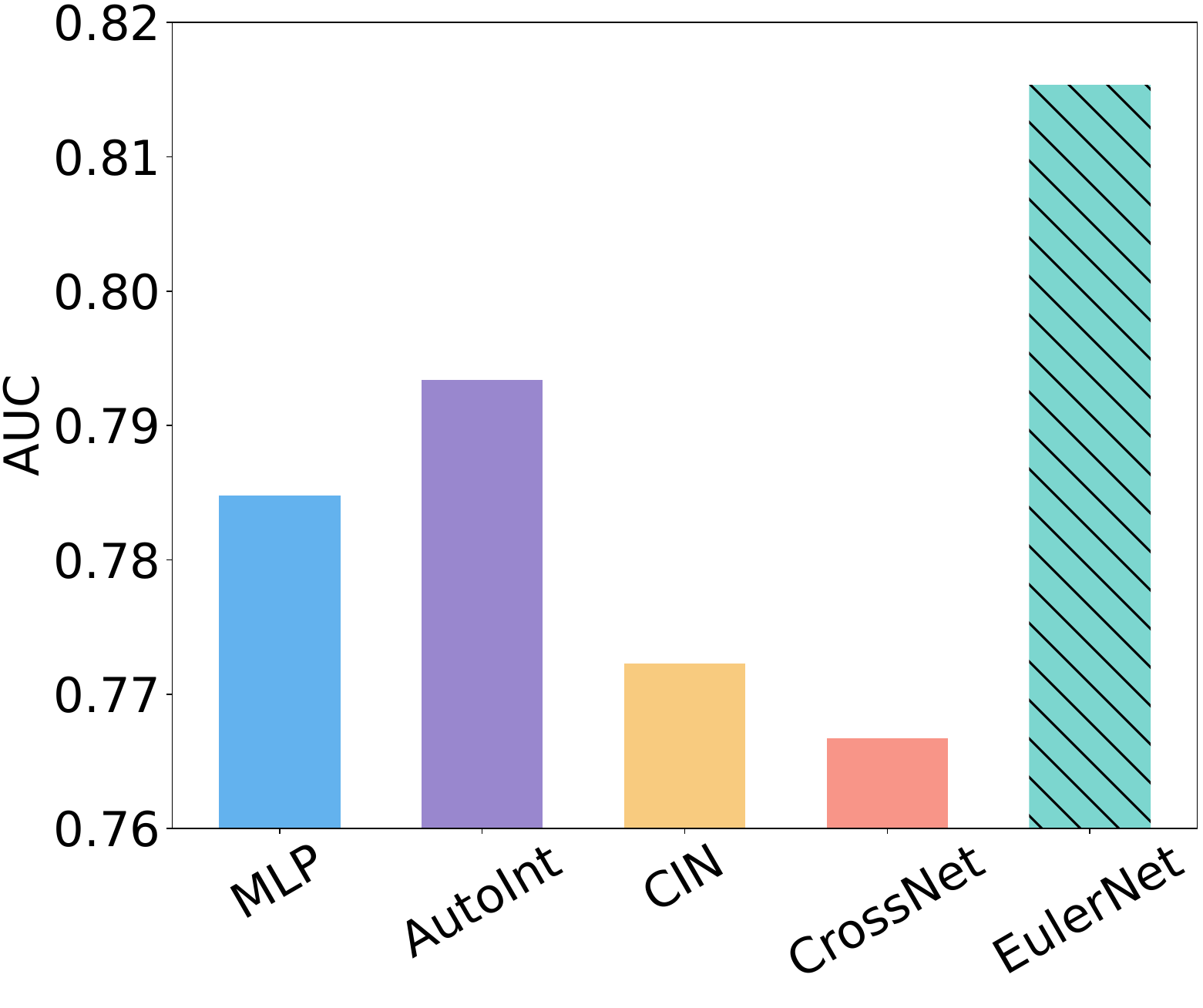}
  }
  \caption{Ablation study of UFIN on the Amazon dataset.}
  \label{fig:alb}
\end{figure}

In this part, we analyze the impact of each proposed technique or component on the model performance.
We propose four variants as: (1) \underline{\emph{w/o KD}} removing the knowledge distillation procedure, (2) \underline{\emph{w/o MoE-E}} removing the MoE model of the encoder ($\bm z = \bm s$ in Eq.~\eqref{eq:ec-moe}), (3) \underline{\emph{w/o UniF}} without generating universal features ($\bm e_j$ in Eq.\eqref{eq:unf}), \ie directly feeding the latent vector $\tilde{\bm z}$ (Eq.~\eqref{eq:unf}) into an MLP, and (4) \underline{\emph{w/o Uid}} without fusing \emph{user\_id} features (See Section~\ref{sec: faf}).
We show the results of ablation study in Figure~\ref{fig:alb}(a).
We observe that all the proposed components are effective to improve the model performance.

Besides, we explore the effect of the feature interaction experts (Section~\ref{sec: fmoe}).
We show the results of five methods (\ie MLP, AutoInt~\cite{song2019autoint}, CIN~\cite{lian2018xdeepfm}, CrossNet~\cite{wang2021dcn}, EulerNet~\cite{tian2023eulernet}) in Figure~\ref{fig:alb}(b).
We can observe that choice of EulerNet achieves the best performance, indicating that adaptive feature interaction learning is the key component to improve the model capacity for domain adaptation.

\subsubsection{Visualizing the Universal Features}\label{exp: case}
\begin{figure}[!h]
    
    {
    \begin{minipage}[t]{1.\linewidth}
        \centering
        \includegraphics[width=1.\textwidth]{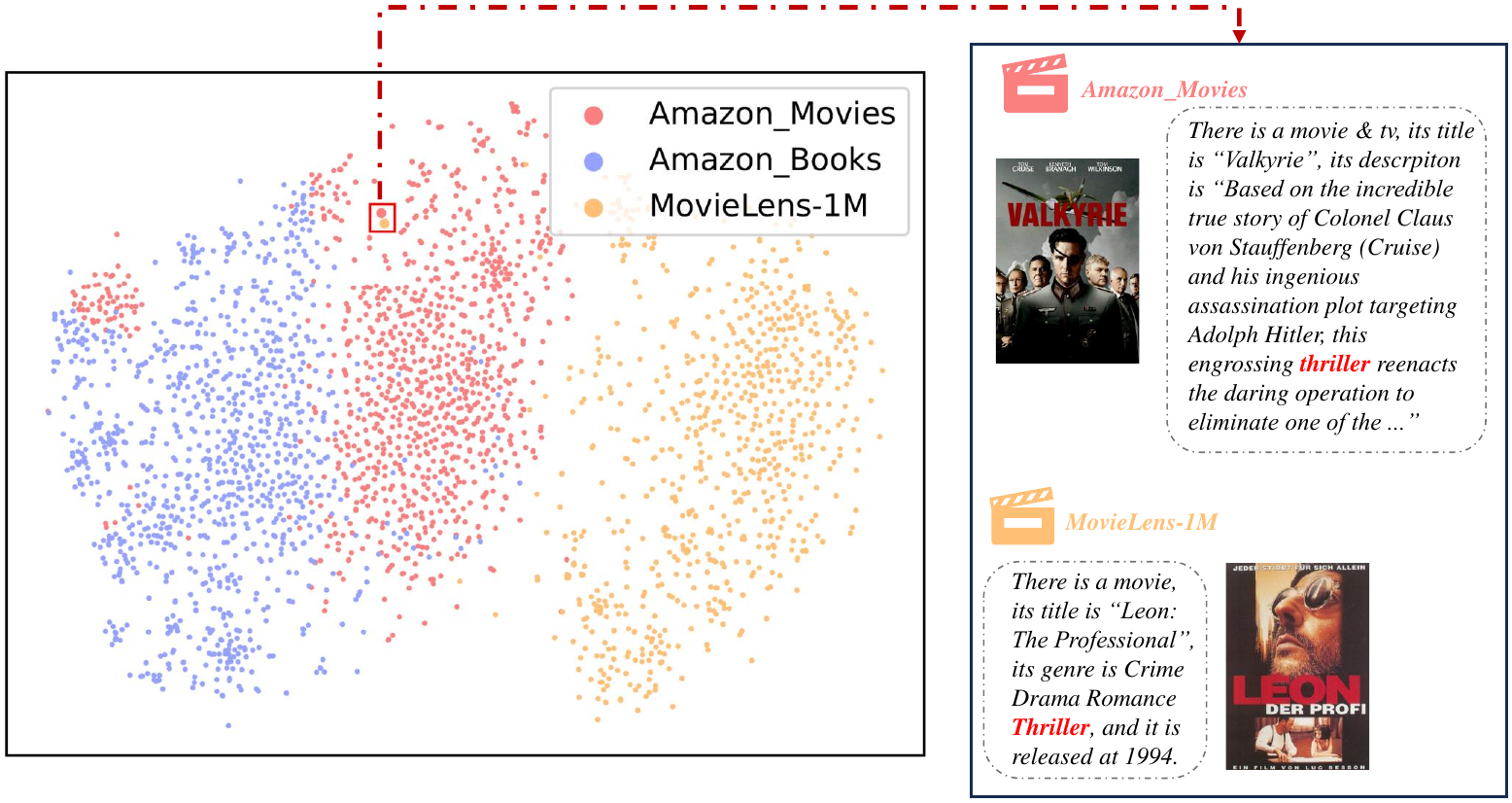}
    \end{minipage}}
    %
    
    \caption{Visualization of the universal features.}
    \label{fig:ts}
\end{figure}

In the literature of CTR predictions, there are few works that can transfer knowledge across different platforms.
To gain a deeper comprehension of the actual knowledge being transmuted, we visually represent the ubiquitous feature representations (See Eq.~\eqref{eq:unf}) on the datasets of Amazon Movies, Amazon Books, and MovieLens-1M by projecting them onto a two-dimensional plane using t-SNE~\cite{van2008visualizing}.
As shown in Figure~\ref{fig:ts}, we can observe that the distribution of  MovieLens-1M is more closely aligned with Amazon\_Movies than Amazon\_Books.
This is reasonable since the contexts of MovieLens-1M and Amazon\_Movies are similar.
Figure~\ref{fig:ts} further presents a case of two closely aligned interactions (highlight in red circle) in the Amazon\_Movies and MovieLens-1M dataset.
Notably, both sentences encompass the keyword \emph{thriller}, thereby demonstrating our approach's ability to capture the collective attributes across disparate scenarios through general textual semantics.

\subsubsection{Visualizing the Universal Feature Interactions}\label{exp: case}
\begin{figure}[!h]
    
    {
    \begin{minipage}[t]{0.3\linewidth}
        \centering
        \includegraphics[width=1\textwidth]{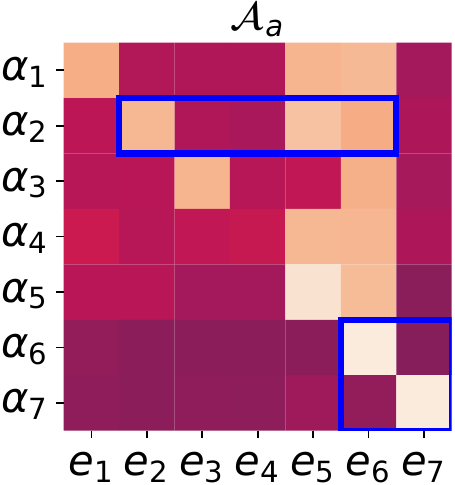}
        \subcaption*{(a) Amazon\_movies} \label{fig:kde_ml1m_ncl}
    \end{minipage}
    \begin{minipage}[t]{0.3\linewidth}
		\centering
		\includegraphics[width=1\textwidth]{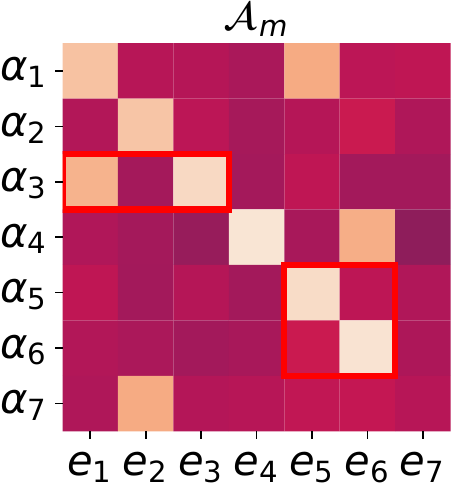}
		\subcaption*{(b) MovieLens} \label{fig:kde_ml1m_lightgcn}
    \end{minipage}
    \begin{minipage}[t]{0.3\linewidth}
        \centering
        \includegraphics[width=1\textwidth]{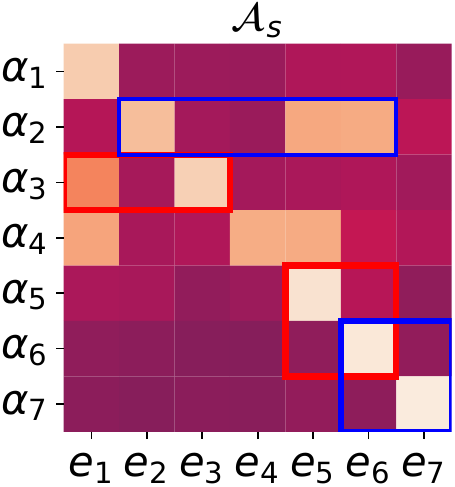}
        \subcaption*{(c) Shared Expert} \label{fig:opp}
    \end{minipage}
    \begin{minipage}[t]{0.065\linewidth}
        \centering
        \includegraphics[width=1\textwidth]{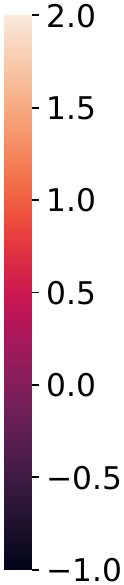}
    \end{minipage}
    \caption{Visualization of the feature interactions.}
    \label{fig:kde}}
\end{figure}

As discussed in Section~\ref{sec:ufi}, UFIN captures the common feature interactions to learn the relatedness between different domains.
To verify the feature interactions learned in UFIN, we visualize the interaction orders (\textit{i.e.}, $\mathcal{A}_j$ in Eq.\eqref{eq:fiexp}) of the Top-1 expert for the Amazon\_movies, MovieLens in Figure~\ref{fig:kde}(a) and (b) respectively.
We can observe that their learned feature interactions exhibit substantial differences.
We further visualize one of their shared experts in Figure~\ref{fig:kde}(c).
Notably, the orders learned in the shared expert exhibit some intersections with those in the domain-specific expert of Amazon\_movies (\textit{i.e.}, $\bm \alpha_2, \bm \alpha_6, \bm \alpha_7$) and MovieLens (\textit{i.e.}, $\bm \alpha_3, \bm \alpha_5, \bm \alpha_6$).
These shared feature interactions enhance the transferability of our model and enable it to capture more generalized collaborative knowledge for CTR predictions.

\subsubsection{Impact of the Selected Experts Number $K$}
\begin{figure}[!h]
  \centering
  
  \subcaptionbox{AUC}{
    \includegraphics[width=0.46\linewidth]{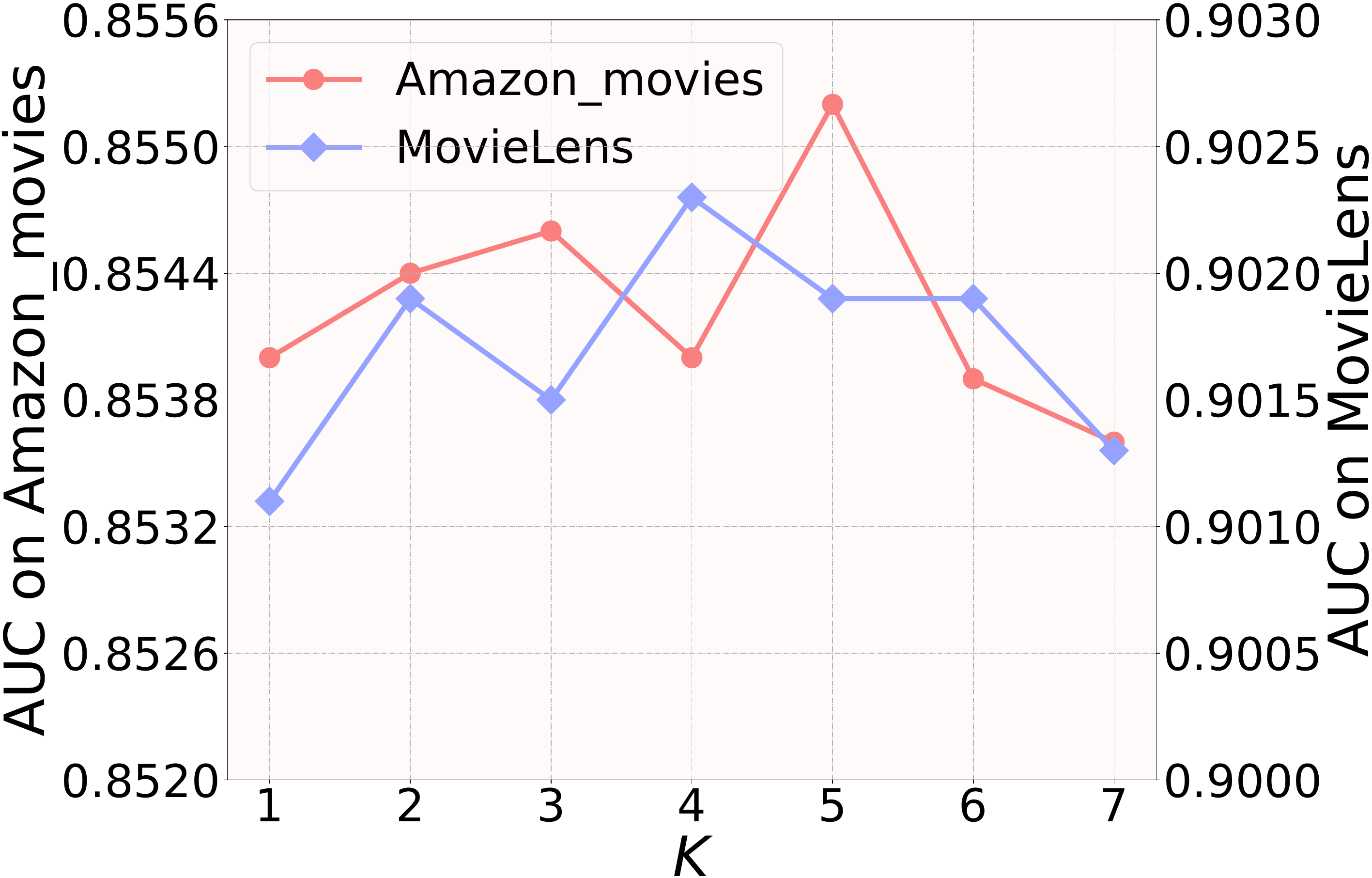}
  }
  \subcaptionbox{LogLoss}{
    \includegraphics[width=0.46\linewidth]{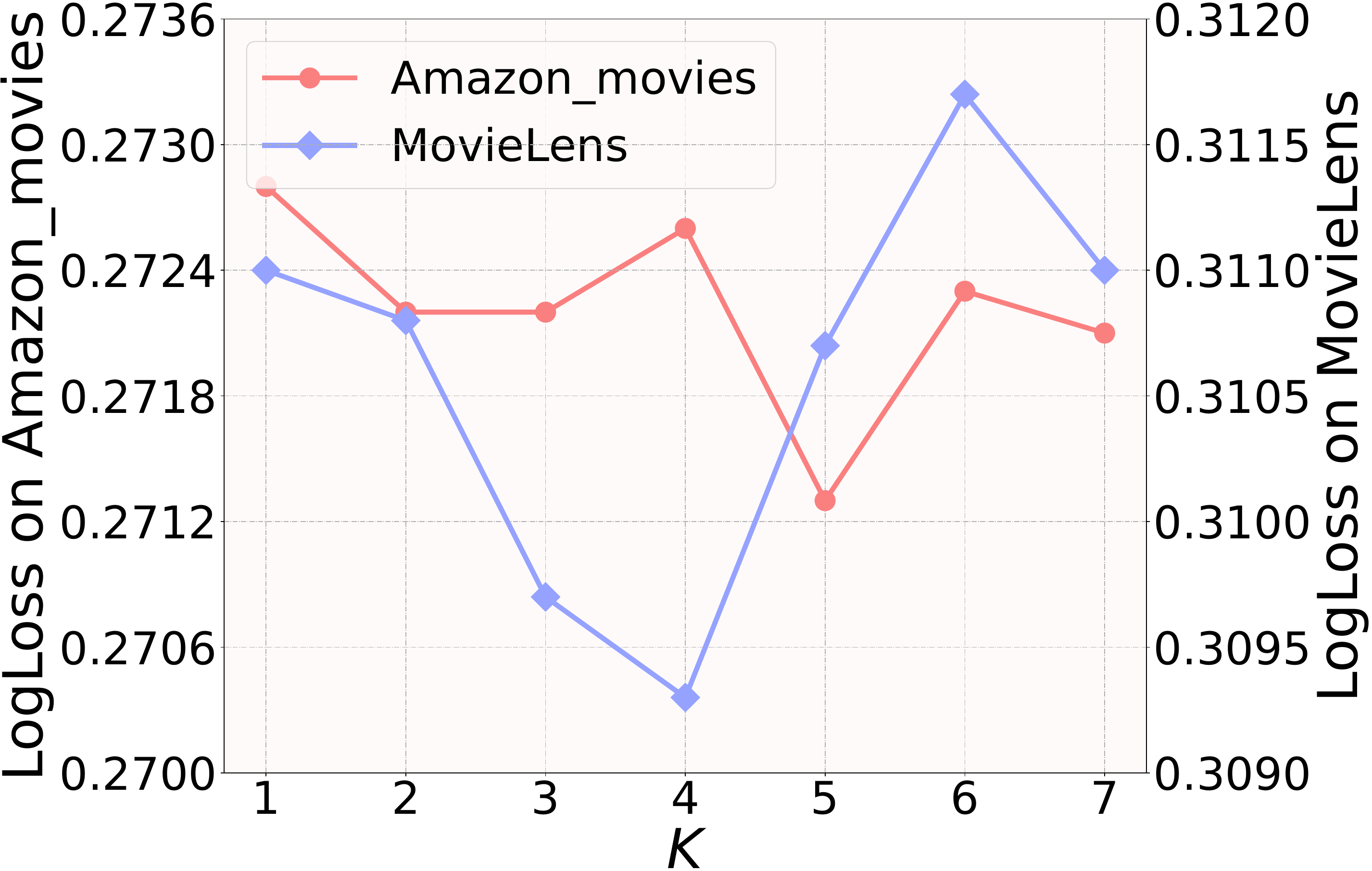}
  }
  
  \caption{Impact of the selected experts number.}
  \label{fig:hyp}
\end{figure}

In the expert selection part, the $K$ value of TopK function (See Eq.\eqref{eq:esele}) can balance the number of domain shared experts.
To analyze the influence of $K$, we vary $K$ in the range of 1 to 7 and report the results in Figure~\ref{fig:hyp}.
We can observe that the performance on the MovieLens dataset increases as $K$ increases from 1 to 4, indicating that the common feature interactions learned in the shared experts are valuable.
Whereas on the Amazon\_Movies dataset, UFIN achieves the best performance as $K$ increases to 5.
However, the model performance decreases when $K$ exceeds 5.
This indicates that adding more shared experts may hinder the learning of domain-specific knowledge that hurt the model performance.

\subsubsection{The analysis of the Prompt Design}\label{sec:prompt}
In this section, we explore the impact of different prompts construction methods on training UFIN. 
Specifically, we consider several rules for constructing
prompts: 
1) Omit auxiliary text descriptions and directly combine feature fields and values using ":";
2) Mask the feature fields with a meaningless
unified word “Field”; 
3) Remove certain feature field;

\begin{quote}
\underline{Base}: There is a user, whose gender is male, and occupation is student. There is a movie,
its title is “Leon:
The Professional”,
its genre is Crime
Drama Romance
Thriller, and it is
released at 1994.

\underline{Prompt-1}: gender: male; occupation: student; title: “Leon:
The Professional”; genre: Crime
Drama Romance
Thriller; release year: 1994.

\underline{Prompt-2}: There is a user, whose Field is male, and Field is student. There is a movie,
its Field is “Leon:
The Professional”,
its Field is Crime
Drama Romance
Thriller, and it is
Field at 1994.

\underline{Prompt-3}: There is a user, whose gender is male, and occupation is student. There is a movie,
its title is “Leon:
The Professional”,
its genre is Crime
Drama Romance
Thriller.
\end{quote}

\begin{figure}[!h]
  \centering
  
  \subcaptionbox{AUC}{
    \includegraphics[width=0.455\linewidth]{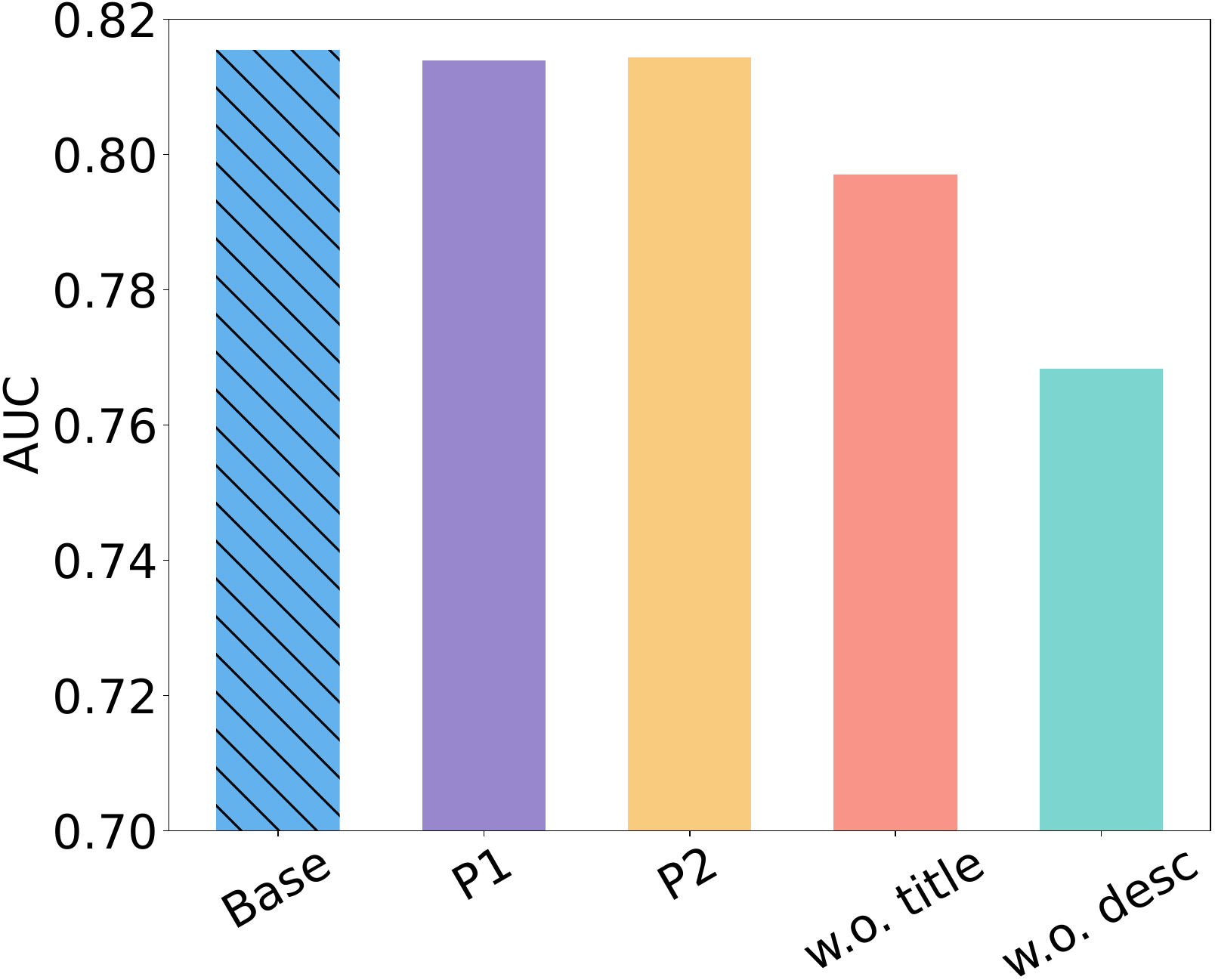}
  }
  \subcaptionbox{LogLoss}{
    \includegraphics[width=0.475\linewidth]{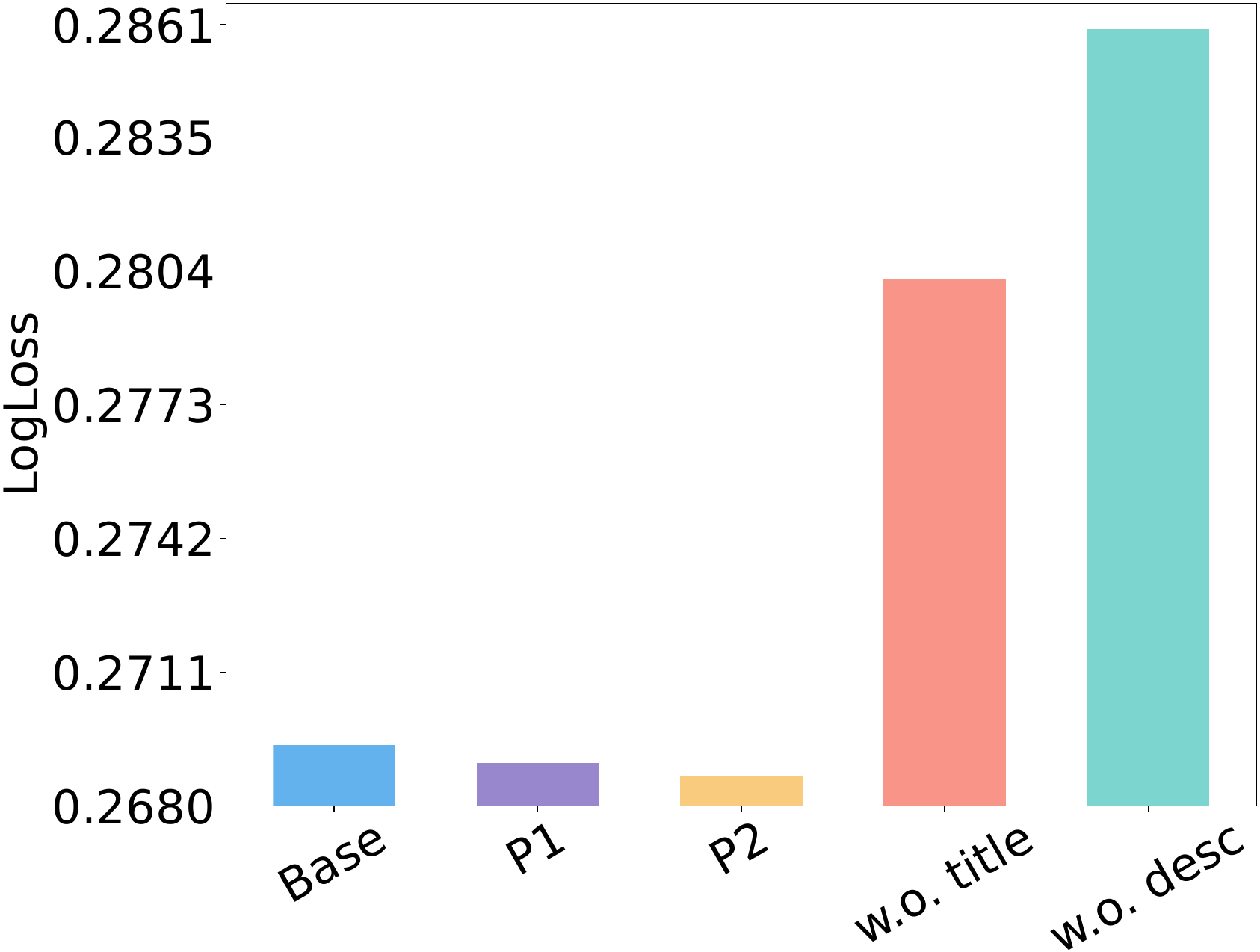}
  }
  
  \caption{Performance on Amazon w.r.t. different prompts.}
  \label{fig:prompt_am}
\end{figure}

\begin{figure}[!h]
  \centering
  
  \subcaptionbox{AUC}{
    \includegraphics[width=0.455\linewidth]{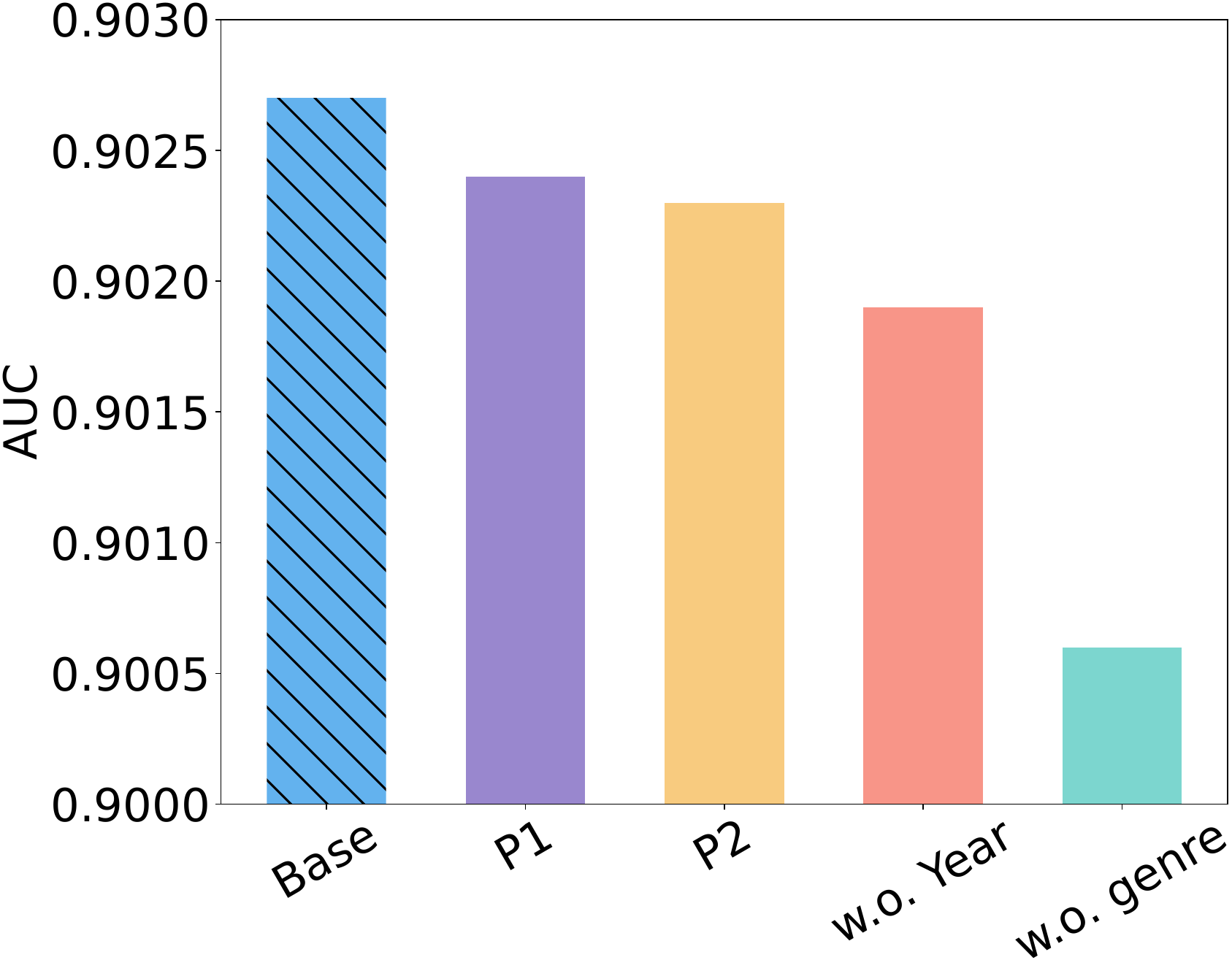}
  }
  \subcaptionbox{LogLoss}{
    \includegraphics[width=0.475\linewidth]{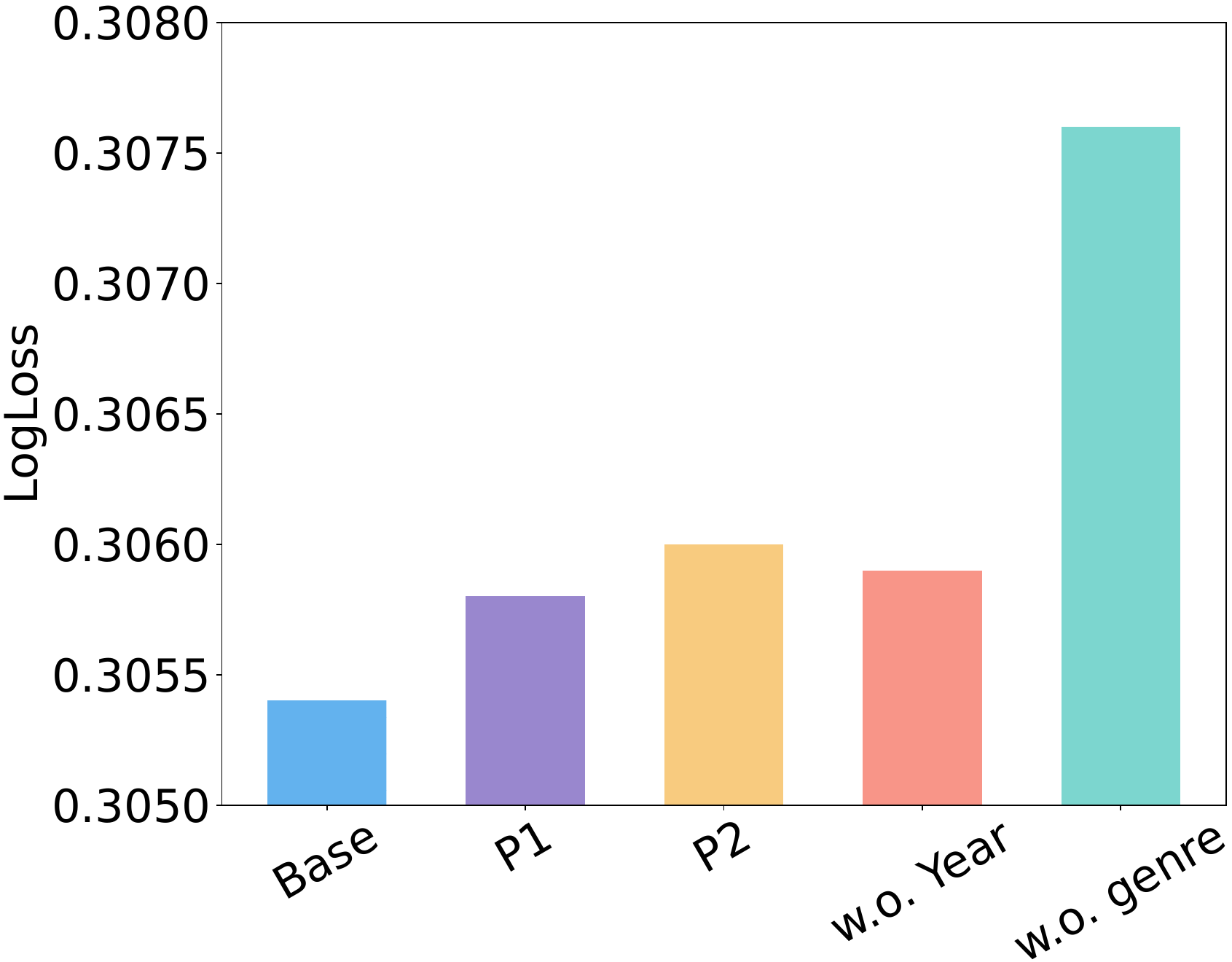}
  }
  
  \caption{Performance on MovieLen-1M w.r.t. different prompts.}
  \label{fig:prompt_mv}
\end{figure}

We utilize three types of prompts to train UFIN, and the results on both the Amazon and MovieLens-1M datasets are depicted in Figures~\ref{fig:prompt_am} and \ref{fig:prompt_mv} correspondingly.
We can observe that on both the Amazon and MovieLens-1M datasets, the performance disparity of the model between Prompt-1, Prompt-2, and the base on both the Amazon and MovieLens-1M datasets is negligible, indicating that our model is not sensitive to the design or format of the template content.
Regarding Prompt-3 on the Amazon dataset, we eliminated the \emph{title} (\underline{w.o. title}) and \emph{description} (\underline{w.o. desc}) fields.	
The analysis from Figure~\ref{fig:prompt_am} reveals that eliminating either the \emph{title} or \emph{description} results in decreased model performance, implying the importance of the semantic information contained within these text features.	
Besides, the removal of the \emph{description} field leads to a greater decline in model performance. This disparity is due to the \emph{description} containing more substantial information compared to the \emph{title}.	
On the MovieLens-1M dataset, we remove the fields \emph{release year} (\ie \underline{w.o. Year}) and \emph{genre} (\ie \underline{w.o. genre}).
As shown in Figure~\ref{fig:prompt_mv}, we can observe that the model's performance changes little after removing \emph{release year}, but there is a sharp decline in the model's performance after removing the \emph{genre} field.
This is because compared to the movie \emph{genre}, \emph{release year} contains fewer personalized item information.
These findings suggest that our approach demonstrates minimal correlation with the design of the prompt text but exhibits a strong relationship with the quantity of information (\emph{personalized features}) it encompasses.	

\section{Related work}

\subsection{Feature Interaction Learning.}
Predicting the probability of users clicking ads or items is critical in  online advertising~\cite{agarwal2014laser} and recommender systems.
The key to achieve a great CTR performance is to learning effective feature interactions.
Typically, most of existing CTR models~\cite{xu2020learning, lu2021dual, pan2018field, blondel2016higher} adopt the embedding look-up and feature interaction learning framework to capture the underlying correlation between users and items.
For example, FM~\cite{rendle2010factorization} introduces low-dimensional vectors to represent the features and uses inner product of them to capture second-order feature interactions.
Besides, FwFM~\cite{pan2018field} and FmFM~\cite{sun2021fm2} propose incorporating field information to improve the pair-wise feature interactions.
However, these approaches only focus on learning the low-order feature interactions, which largely degrades the model capacity and results in the suboptimal performance.

Recently, a number of approaches~\cite{lian2018xdeepfm, wang2021dcn} are proposed to model the high-order feature interactions.
Among them, xDeepFM~\cite{lian2018xdeepfm} conducts multi-layer convolution operations on the concatenated vectors to model high-order interactions; DCNV2~\cite{wang2021dcn} performs the kernel product on the concatenated feature vectors to construct higher-order feature interactions.
These approaches have largely raised the performance bar of CTR predictions.
Typically, these approaches follow a common paradigm to model the feature interactions: they
first pre-define a maximal order and only consider conducting feature interactions within the pre-defined orders.
Though effective to some extend, the orders of feature interactions are empirically designed, which cannot accurately capture the underlying feature relationship in the real-world scenarios.
Further, due to the exponential growth of high-order feature combinations, such approaches often set a small order, which cannot scale to the high-order cases in the industrial scenarios.

Considering the above limitations, some approaches~\cite{tian2023eulernet, cheng2020adaptive, cai2021arm} are proposed to automatically learn the feature interaction orders from the data.
The core idea of these approaches is to encode the features into a special vector space (\eg logarithm vector space).
As such, the complicated feature interactions are converted to the simple linear combinations and the orders are cast into the learnable coefficients, enabling the adaptive learning of the interaction orders.
However, previous studies mainly focus on learning the orders within a single-domain, and seldom investigate the transferability of the order information across different domains or tasks.

\subsection{Multi-Domain CTR predictions.}
Traditional CTR prediction models are developed for modeling the user's interest within a single domain, \ie they are training using examples collected from a single domain and serving the prediction of a single task.
However, in large-scale real-world industrial platforms, the user behavior data are often collected from multi-domains.
Merely mixing all the data and training
a single shared CTR model cannot yield satisfactory results
across all domains owing to the substantial distribution variance
among diverse scenarios, which is called domain seesaw phenomenon in recommender systems.
Such paradigm severely restricts the efficient utilization of extensive user behavior data in business scenarios.

To address these issues, a number of approaches~\cite{sheng2021one, chang2023pepnet, zhang2022scenario, zhou2023hinet} are proposed to enhance the information sharing and improve the multi-domain performance.
Among them, SharedBottom~\cite{ruder2017overview} introduces a unique embedding layer to learn a commonly shared feature representations for different domains.
Besides, MMoE~\cite{ma2018modeling} extends the multi-task learning (MTL) methods into multi-domain scenarios, by regarding each domain as a specific task; HMoE~\cite{tang2020progressive} extends MMoE to scenario-aware experts using a gradient cutting trick to explicitly encode scenario correlations.
Despite the progress, these approaches cannot effectively exploit the domain relationship and suffer from the degradation of model capability.

As a promising research direction, some recent studies~\cite{tang2020progressive, zou2022automatic, sheng2021one, chang2023pepnet} propose to explicitly introduce the domain-shared parameters to compactly learn the common knowledge across different domains.
The core idea of these approaches is
to introduce a shared neural network for learning the common
knowledge across diverse domains, while simultaneously
integrating multiple domain-specific sub-networks to capture
the distinct characteristics of each domain.
For example, STAR~\cite{sheng2021one} proposes a star topology architecture, which consists of a shared center network and multiple domain-specific networks for the adaptation of multi-domain distributions;
PEPNet~\cite{chang2023pepnet} introduces an embedding personalized network to align multi-domain feature representations.
Despite the progress, these approaches still rely on explicit embedding look-up operation for the given feature IDs, which impairs the inherent semantic of features and makes it difficult to be applied in different platforms.





\subsection{Semantic CTR Prediction Models.}
With the development of natural language processing (NLP), the large language models (LLMs~\cite{zhao2023survey}) have shown excellent language modeling capacity in various downstream tasks, which promises researchers to apply LLMs into the recommendation tasks.
In the context of CTR predictions, the large language models (LLMs) have two common application methods in CTR prediction models:  scoring/ranking function and feature engineering encoder.

For the first kind of application, the LLMs mainly serve as the learning backbone to generate predictive results.
For example, P5~\cite{geng2022recommendation} converts different recommendation task into text generation and employs T5~\cite{raffel2020exploring} model to generate the result.
Besides, M6-Rec~\cite{cui2022m6} uses the M6~\cite{lin2021m6} model to deliver the recommendation in a prompting learning framework.
These approaches exploit the textual data as a general input form, which can effectively bridge the semantic gap between different domains, enabling them to be effectively transferred to a new domain or platform.

Despite the progress, compared to the traditional CTR prediction models, LLMs cannot effectively capture the collaborative patterns that severely limits the model capacity.
As a promising approach, a recent study~\cite{li2023ctrl} proposes a contrastive learning framework to align the knowledge between LLMs and collaborative models.
Due to the limited capacity of collaborative model, it cannot adequately learn the common knowledge across different domains.
In contrast, our solution is to utilize the LLMs to generate universal features from textual data, learning the collective attributes of different domains.
As such, we can learn the universal feature interactions to capture the generalized collaborative patterns across diverse domains, naturally integrating the world knowledge of LLMs and the collaborative knowledge between users and items.

\section{Conclusion}
In this paper, we propose the Universal Feature Interaction Network  (\textbf{UFIN}) for multi-domain CTR prediction.
Unlike previous approaches that heavily rely on modeling ID features for developing the CTR predictions, our approach leverage the textual data to learn the universal feature interactions.
Specifically, we regard the text and features as two modalities that can be mutually converted.
As such, we employ a LLM-based encoder-decoder architecture to transform the data from text modality to feature modality, obtaining universal feature representations.
Building upon this foundation, we design an adaptive feature interaction model enhanced by a mixture-of-experts (MoE) architecture for capturing the generlized feature interactions across different domains.
To effectively learn the collaborative patterns across different domains, we propose a multi-domain knowledge distillation framework to improve the training of our approach.

As future work, we will explore how to better integrate anonymous features with generated universal feature representations.
In addition, we will also consider incorporating the Conversion Rate (CVR) prediction task into our approach to capture more effective correlation between different tasks.

\section{Proofs of Theorem~\ref{thee}}\label{sec:pro}
\begin{proof}

    By contradiction, we assume there exists two domains $\mathcal{D}_u$ and $\mathcal{D}_v$ ($u \neq v$) that satisfies $\mathcal{S}_u \cap \mathcal{S}_v = \emptyset$.

    According to the principle of inclusion-exclusion, we have:

    \begin{align*}
            |\mathcal{S}_u \cup \mathcal{S}_v | &= |\mathcal{S}_u| + |\mathcal{S}_v| - |\mathcal{S}_u \cap \mathcal{S}_v| \\& = K + K - 0 \\ & = 2K > 2 \times \lceil{L/2}\rceil \geq L
    \end{align*}

    On the other hand, since $\mathcal{S}_u \subseteq \mathcal{A}$ and $\mathcal{S}_v \subseteq \mathcal{A}$, we have
    $\mathcal{S}_u \cup \mathcal{S}_v \subseteq \mathcal{A}$.

    Therefore, we have:  
    \begin{align*}
        |\mathcal{S}_u \cup \mathcal{S}_v | \leq |\mathcal{A}| = L.
    \end{align*}
    
    Since the assumption leads to a contradiction, our initial assumption that $\exists u \neq v, \mathcal{S}_u \cap \mathcal{S}_v = \emptyset$ must be \textbf{False}.

    Therefore,  $\forall u \neq v, \mathcal{S}_u \cap \mathcal{S}_v \neq \emptyset$.
\end{proof}

\bibliographystyle{IEEEtran}
\bibliography{IEEEabrv,reference}
\balance

\end{document}